\documentclass[prd,twocolumn,superscriptaddress,amsfonts,amssymb,amsmath,showpacs]{revtex4-2}
\usepackage{bm}
\usepackage{latexsym}
\usepackage[latin1]{inputenc}
\usepackage{graphicx}
\usepackage{amsmath}
\usepackage{amssymb}
\usepackage{amsthm}
\usepackage{relsize}
\usepackage{palatino}
\usepackage{mathpazo}
\usepackage{textcomp}
\usepackage{float}
\usepackage{booktabs}
\usepackage{dcolumn}
\usepackage{booktabs}
\usepackage{multirow}
\usepackage{tikz}
\usepackage{hyperref}
\hypersetup{
    colorlinks=true,
    linkcolor=purple,
    filecolor=magenta,      
    citecolor=blue
}
\usepackage{amsmath}
\usepackage{xcolor}
\usepackage{orcidlink}
\usepackage[caption=false]{subfig}
\usepackage{commath}
\makeatletter
\long\def\dddddot#1{%
  {\mathop {#1}\limits ^{\vbox to-1.4\ex@ {\kern -\tw@ \ex@ \hbox {\normalfont .....}\vss }}}%
}
\long\def\multidots#1#2{%
  \count@=0
  {{\mathop {#2}\limits ^{\vbox to-1.4\ex@ {\kern -\tw@ \ex@ \hbox {\normalfont %
  \loop%
  \ifnum#1>\count@%
  .%
  \advance\count@ by1%
  \repeat%
  }\vss }}}}%
}
\makeatother

\begin{document}

\title{\bf  Tracing cosmic evolution through Weyl-Type $f(Q,T)$ gravity model: theoretical analysis and observational validation}

\author{Rahul Bhagat\orcidlink{0009-0001-9783-9317}}
\email{rahulbhagat0994@gmail.com}
\affiliation{Department of Mathematics, Birla Institute of Technology and
Science-Pilani, Hyderabad Campus, Hyderabad-500078, India.}
\author{Francisco Tello-Ortiz \orcidlink{0000-0002-7104-5746}}
\email{francisco.tello@ufrontera.cl}
\affiliation{Departamento de Ciencias Fisicas, Universidad de La Frontera, Casilla 54-D, 4811186 Temuco, Chile.}

\author{B. Mishra\orcidlink{0000-0001-5527-3565}}
\email{bivu@hyderabad.bits-pilani.ac.in}
\affiliation{Department of Mathematics, Birla Institute of Technology and
Science-Pilani, Hyderabad Campus, Hyderabad-500078, India.}

\begin{abstract}
We investigate the cosmic evolution of the Universe across different cosmological epochs in exponential Weyl-type \( f(Q, T) \) gravity model. The theoretical analysis involves a detailed dynamical system approach, where we define dimensionless variables and derive a system of linear differential equations to identify critical points corresponding to the radiation, matter and de Siter phase. The findings show the transition from deceleration to acceleration phase, with stable and unstable critical points characterizing different phases of the evolution. In the second approach, we validate the theoretical predictions by using observational data from Cosmic Chronometers ($CC$) and $Pantheon^+$ datasets. We constrain the Hubble parameter and subsequently analysed the other cosmological and geometrical parameters. In this approach also, the transition from deceleration to acceleration has been confirmed, with the equation of state (EoS) parameter approaching $\Lambda$CDM at late times. The further validate this, we present the behaviour of state finder pair. We obtain the age of the Universe $13.81$ Gyr according to $CC$ data and $13.96$ Gyr with the $Pantheon^+$ dataset. The model behaviour in both the approaches shows strong agreement in the late-time behavior of the Universe. The evolutionary behaviour of Hubble parameter and distance modulus, reinforcing the reliability of the Weyl-type $f(Q, T)$ gravity model in describing the expansion history of Universe. \\

\end{abstract}

\maketitle

\section{Introduction} 
The recent cosmological observations indicate the accelerated expanding behaviour of the Universe \cite{Riess_1998_116,Perlmutter_1998_517}. The simplest explanation for this phenomenon is the cosmological constant ($\Lambda$). However, the potential for a dynamic nature has led to two main approaches to its explanation. The first approach adheres to general relativity (GR) and introduces the idea of an exotic energy known as dark energy (DE). The second approach attributes the new degrees of freedom to modify in the gravitational interaction leading to extended theories and including general relativity as a limit \cite{Weinberg_1989_61}. The challenges $\Lambda$CDM model have prompted a reevaluation of alternative ideas for different cosmological models \cite{Appleby_2018_07}. Focusing on the early and late Universe paradigm derived from the cosmic tensions problem, several models have individually modified standard cosmology in distinct sectors \cite{DiValentino_2021_38,Abdalla_2022_34,Riess_2021_908}. Solutions for the late-time regime of the Universe involve altering the recent cosmic evolution without changing the early Universe \cite{Wong_2019_498}. These solutions include decaying models of dark matter or DE, running vacua models \cite{Gibbons_1987}, and sign-switching models \cite{Akarsu_2021}. Unless one consider fundamental changes to the underlying principles of standard cosmology, the combination of Type Ia supernovae (SNIa) \cite{Riess_2004_607}, $CC$ \cite{Moresco_2022_25}, and baryonic acoustic oscillation (BAO) \cite{Raichoor_2020_500} data significantly constrains possible deviations from the $\Lambda$CDM model. Whereas, the early Universe proposals primarily focus on modifications that reduce the sound horizon, which implies a higher Hubble constant for a fixed CMB angular scale. There are some examples that include early Universe DE models and more exotic neutrino models \cite{Simon_2005_71}. However, the necessary reduction in the sound horizon makes these models incompatible with the growth of large-scale structure \cite{Perenon_2015_11}. This leads to the global rescaling of the Hubble evolution of the Universe. This can be achieved by modifying GR, which leads to the modified theories of gravity.

The formulation of GR is based on Riemann geometry, where the curvature is derived from the metric. One of the geometrical modifications is the nonmetricity based gravitational theory, known as symmetric teleparallel gravity (STG), in which the curvature in GR is replaced with the  Levi-Civita connection $\Gamma$ \cite{Jimenez_2018_039}. However, using the coincident gauge, it may conflict with the coordinate system \cite{Zhao_2022_82}. So, to overcome this the $f(Q)$ gravity theory in covariant form has been developed which enable the identification of a suitable non-vanishing affine connection for a given metric \cite{Harko_2018_98}, $Q$ being the nonmetricity scalar. Several cosmological models in this gravity are available in the literature pertaining to the issue of  late time cosmic dynamics \cite{Lazkoz_2019_100,Jimenez_2020_101,Barros_2020_30, Anagnostopoulos_2021_822}. The $f(Q)$ gravity further extended to $f(Q,T)$ gravity with the inclusion of the trace of energy momentum tensor\cite{Xu_2020_80}. Zia et al. \cite{Zia_2021} introduced a transitional cosmological model consistent with the observational deceleration parameter within \( f(Q, T) \) gravity. Sharif et al. \cite{Sharif_2024} examined the reconstruction criteria within the \( f(Q, T) \) gravity framework using the ghost DE model. Narawade et al. \cite{Narawade_2023_992} constrained an accelerating cosmological model in \( f(Q, T) \) gravity and performed a dynamical system analysis. Pati et al.\cite{Pati_2021} developed a quintessence model in the context of a hybrid scale factor. Najera and Fajardo \cite{Najera_2021_34} tested five \( f(Q, T) \) models, demonstrating that these models reduce to the $\Lambda$CDM model with specific parameter values. Additional research on \( f(Q, T) \) gravity is also available in the literature \cite{Bhagat_2023_42,Narawade_2024Baryon}.

 In this geometric framework, the magnitude of the vector $\omega_{\mu}$ governs the entirety of nonmetricity in Weyl geometry \cite{Xu_2020_80,Wheeler_2014}. To capture a comprehensive dynamical description of the gravitational field, we need to integrate two terms associated with the energy and mass of the vector field into the gravitational action. With the gravitational Lagrangian and geometric action in place, the gravitational field equations through standard variational methods can be obtained. The resulting equations capture gravitational phenomena in Weyl geometry with a massive vector field coupled to the matter energy-momentum tensor within a globally flat geometry \cite{Alvarez_2017}. By scrutinizing these field equations, the divergence of the matter energy-momentum tensor does not vanish within this approach to gravitational interaction \cite{Gomes_2019}. Haghani et al. \cite{Haghani_2013_88} have explored an extension of the Weyl$-$Cartan$-$Weitzenb$\ddot{o}$ck (WCW) and teleparallel gravity, where the Weitzenb$\ddot{o}$ck condition ensures the exact cancellation of curvature and torsion in a Weyl$-$Cartan geometry is incorporated into the gravitational action through a Lagrange multiplier. Yang et al. \cite{Yang_2021_81} have examined the geodesic deviation equation in Weyl$-$type \( f(Q, T) \) gravity. Bhagat et al.\cite{Bhagat_2023_41} have investigated the cosmological features of Weyl$-$type \( f(Q, T) \) gravity by utilizing the parameterized form of the Hubble parameter. Some more research on Weyl type $f(Q,T)$ gravity can be seen in Refs. \cite{Zhadyranova_2024,Bhagat_ASPdyna2024}.

This paper is organized as follows: In Section \ref{Sec:2}, we present the framework for the field equations of Weyl-type $f(Q,T)$ gravity. In Section \ref{Sec:3}, we discuss the dynamical system analysis with the dimensionless variables and elaborate the evolutionary behavior of the Universe through the critical points. In Section \ref{Sec:4}, we incorporate the cosmological observational data such as, $CC$ and $Pantheon^+$ to constrain the dynamical and geometrical parameters. In Section \ref{Sec:5}, we present the summary and conclusion of the cosmological model. 

\section{Weyl type $f(Q,T)$ gravity}\label{Sec:2}

The action for Weyl type $f(Q,T)$ \cite{Xu_2020_80} is,
\begin{eqnarray}\label{eq:1}
S &=& \int \sqrt{-g} \big[\frac{1}{16\pi G} f(Q,T) -\frac{1}{4} W_{\mu\nu} W^{\mu\nu}-\frac{1}{2}m^{2} \omega_{\mu} \omega^{\mu}\nonumber\\&&
+\lambda (R +6\bigtriangledown_\alpha \omega^ \alpha-6 \omega_\alpha \omega^ \alpha) +\mathcal{L}_m \big]dx^4 ~,
\end{eqnarray}
 where $W_{\mu\nu}=\triangledown_\nu \omega_\mu-\triangledown_\mu \omega_\nu$ is the form of Weyl vector, $\mathcal{L}_m$ be the matter Lagrangian with $m$ be the mass. The second term $\frac{1}{4} W_{\mu\nu} W^{\mu\nu}$  and third term $\frac{1}{2}m^{2} \omega_{\mu} \omega^{\mu}$ respectively represents the standard kinetic term and mass term of the vector field. The semi-metric connection in Weyl geometry is,
\begin{equation}\label{eq:2}
     \tilde \Gamma ^\lambda_{\mu\nu}=\Gamma^{\lambda}_{\mu\nu}+g_{\mu\nu}w^\lambda-\delta_\mu^\lambda w_\nu - \delta_\nu^\lambda w_\mu,
\end{equation}
where $\Gamma ^\lambda_{\mu\nu}$ represents the Christoffel symbol, which is to be constructed with respect to the metric $g_{\mu\nu}$. The nonmetricity tensor is,
\begin{align}\label{eq:3}
 Q_{\alpha\mu\nu}&\equiv\tilde\triangledown_\alpha g_{\mu\nu}=\partial_\alpha g_{\mu\nu} -\tilde \Gamma ^\rho _{\alpha \mu}g_{\rho \nu}-\tilde \Gamma ^\rho_{\alpha\nu}g_{\rho\mu}\nonumber\\& = 2\omega_\alpha g_{\mu\nu}.
\end{align}
and the scalar nonmetricity becomes
\begin{equation}\label{eq:4}
Q \equiv -g^{\mu\nu} (L^\alpha_{\beta\mu}L^\beta _{\nu\alpha} - L^\alpha _{\beta\alpha} L^\beta _{\mu \nu}),
\end{equation}
with
\begin{equation}\label{eq:5}
L^\lambda _{\mu\nu}=-\frac{1}{2}g^{~\lambda\gamma}(Q_{~\mu\gamma\nu}+Q_{~\nu\gamma\mu}-Q_{~\gamma\mu\nu}).
\end{equation}
Substituting Eq. \eqref{eq:3} in Eq. \eqref{eq:5}, one can obtain
\begin{equation}\label{eq:6}
Q=-6 \omega ^2.
\end{equation}
By varying the action with respect to the vector field, the generalized Proca equation becomes
\begin{equation}\label{eq:7}
\bigtriangledown^{\nu}W_{\mu\nu}-(m^2+12\kappa^2~f_Q+12\lambda)\omega_\mu=6\bigtriangledown_{\mu}\lambda.
\end{equation}
Applying the variation principle to the action, the field equation of Weyl type $f(Q, T)$ gravity obtained as,
 \begin{multline}\label{eq:8}
\frac{1}{2}(T_{\mu\nu}+S_{\mu\nu})-\kappa^2 f_T (T_{\mu\nu}+\Theta_{\mu\nu})=-\frac{\kappa^2}{2}g_{\mu\nu}f\\-6\kappa^2 f_Q \omega_\mu \omega_\nu+\lambda(R_{\mu\nu}-6\omega_\mu \omega_\nu+3g_{\mu\nu}\triangledown_\rho \omega^\rho)\\+3g_{\mu\nu}\omega^\rho\triangledown_\rho\lambda-6\omega_{(\mu}Q_{\nu)}\lambda+g_{\mu\nu}\square \lambda- \triangledown_\mu\triangledown_\nu \lambda,
\end{multline}
where the derivatives of $f(Q, T)$ with respect to $Q$ and $T$ are respectively denoted as $f_Q$ and $f_T$. The energy momentum tensor
\begin{equation}\label{eq:9}
T_{\mu\nu}\equiv -\frac{2}{\sqrt{-g}}\frac{\delta(\sqrt{-g}\mathcal{L}_m)}{\delta g^{\mu\nu}},
\end{equation}
\begin{equation}\label{eq:10}
\Theta_{\mu\nu} \equiv g^{\alpha\beta}\frac{\delta T_{\alpha \beta}}{\delta g_{\mu\nu}}=g_{\mu\nu}\mathcal{L}_m-2T_{\mu\nu}-2g^{\alpha\beta}\frac{\delta^2\mathcal{L}_m}{\delta g^{\mu\nu}\delta g^{\alpha\beta}}.
\end{equation}\\
The rescaled energy momentum tensor $S_{\mu\nu}$ becomes
\begin{multline}\label{eq:11}
S_{\mu\nu}=-\frac{1}{4}g_{\mu\nu}W_{\rho\sigma}W^{\rho\sigma}+W_{\mu\rho}W^\rho_\nu\\-\frac{1}{2}m^2 g_{\mu\nu}\omega_\rho \omega^\rho+m^2\omega_\mu \omega_\nu~.
\end{multline}
To frame cosmological model of the Universe, we consider flat FLRW space-time as
\begin{equation}\label{eq:12}
ds^2=-dt^2+a^2(t)[dx^2+dy^2+dz^2],
\end{equation}
where $a(t)$ is the uniform spatial expansion function. The Hubble parameter, $H = \frac{\dot a}{a}$, where an over dot represents the time derivative. Given the spatial symmetry, the vector field takes the form
\begin{equation}\label{eq:13}
\omega_\nu=[\psi(t),~0,~0,~0]
\end{equation}
and $\omega^2 = \omega_\nu \omega^\nu = -\psi^2(t)$, $Q = 6 \psi^2(t)$. Using the comoving coordinate system, we have $u^\mu = (-1,~0,~0,~0)$, hence $u^\mu \triangledown_\mu = \frac{d}{dt}$. Assuming the Lagrangian of a perfect fluid $\mathcal{L}_m = p$, we get $T_\nu^\mu = diag(-\rho,~p,~p,~p)$ and $\Theta^\mu_\nu = diag(2 \rho + p, -p, -p, -p)$. The generalized Proca equation describing the evolution of the Weyl vector can be expressed as,
\begin{eqnarray}\label{eq:14}
    \dot{\psi} &=& \dot{H} + 2H^{2} + \psi^{2} -3H\psi~,\nonumber\\
    \dot{\lambda} &=& -\frac{1}{6}~m^{2}_{eff}~\psi ~,\nonumber\\
    \partial_{i}\lambda &=& 0~,
\end{eqnarray}
where $m^2_{eff} = m^2 + 12 \kappa^2 f_Q + 12 \lambda$ is the effective dynamical mass of the vector field and $\kappa^2 = \frac{1}{16\pi G}$ is the gravitational coupling constant. Now, the generalised Friedmann equations of Weyl type $f(Q, T)$ gravity derived can be obtained as,

\begin{multline}\label{eq:15}
\kappa^2f_T(\rho+p)+\frac{1}{2}\rho=\frac{\kappa^2}{2}f-\left(6\kappa^2f_Q+\frac{1}{4}m^2\right)\psi^2 \\-3\lambda(\psi^2-H^2)-3\dot\lambda(\psi-H)
\end{multline}
\begin{multline}\label{eq:16}
-\frac{1}{2}p=\frac{\kappa^2}{2}f+\frac{m^2\psi^2}{4}+\lambda(3\psi^2+3H^2+2\dot H)\\ +(3\psi+2H)\dot\lambda+\ddot\lambda   \end{multline}
Using Eq. \eqref{eq:14}, Eq. \eqref{eq:15} and Eq. \eqref{eq:16} can be respectively reduces to, 
\begin{multline}\label{eq:17}
\frac{1}{2}\left(1+2\kappa^2 f_T\right)\rho+\kappa^2f_T p=\frac{\kappa^2}{2}f+\frac{m^2\psi^2}{4}\\+3\lambda(H^2+\psi^2)-\frac{1}{2}m^{2}_{eff}H\psi,
\end{multline}
\begin{multline}\label{eq:18}
    \frac{1}{2}\left(1+2\kappa^2f_T\right)(\rho+p)=\frac{m^{2}_{eff}}{6}\left(\dot{\psi}+\psi^2-H\psi \right)\\+2\kappa^2\dot{f_Q}\psi-2\lambda\dot{H}
\end{multline}
Further substituting $\dot{\psi}$ in Eqn. \eqref{eq:18}, we obtain 
\begin{multline}\label{eq:19}
    \frac{1}{2}\left(1+2\kappa^2f_T\right)(\rho+p)=-2\lambda \Bigl(1-\frac{m^{2}_{eff}}{12\lambda}\Bigl)\dot{H}\\+\frac{m^{2}_{eff}}{3}\left(H^2+\psi^2-2H\psi \right)+2\kappa^2\dot{f_Q}\psi.
\end{multline}
The equation for the energy balance is,
\begin{equation}\label{eq:20}
    \dot{\rho}+3H(\rho+p)=\frac{1}{1+2\kappa^2f_{T}}(2\kappa^2(\rho+p)\dot{f_T}-f_T(\dot{\rho}-\dot{p})).
\end{equation}
After presenting the field equations for Weyl type $f(Q,T)$ gravity, we shall first analyse the dynamical system approach to analysis to understand the evolutionary history of the Universe. So, in the next section, we shall present the dynamical system analysis of the model.
\section{The Dynamical system Analysis}\label{Sec:3}
Dynamical system analysis is an excellent tool to understand and explore the evolutionary history of the Universe in details. In this approach, appropriately choosing the dimensionless variables, the complex cosmological equations can be transformed into simple ordinary differential equations. Then one can obtain the critical points and through the behaviour of these points, the evolutionary history can be studied. The phase portrait provide some additional information on the critical points. In the nonmetricity gravitational theory, Paliathanasis \cite{Paliathanasis_2023_41} studied the evolution of the physical variables in $f(Q)$ gravity for two families of symmetric connections in the spatially flat FLRW geometry. Through a model-independent set of variables, Bohmer et al. \cite{Bohmer_2016_book_dyna} studied a dynamical systems approach of $f(Q)$ gravity models. Pati et al.\cite{Pati_2023_83} derived the dynamic parameters in the general form of higher power nonmetricity to study the evolution of the Universe. Additionally, various prominent works have been done in modified theories of gravity\cite{Lohakare_2023_39,Lu_2019_79,Duchaniya_2023_83}. Now, the generalized Friedmann Eq. \eqref{eq:17} and Eq. \eqref{eq:18} can be reformulated in an effective form as,
\begin{eqnarray}
3H^2&=&\frac{1}{2\lambda}(\rho+\rho_{eff}),\label{eq:21}\\
3H^2+2\dot{H}&=&-\frac{1}{2\lambda}(p+p_{eff}), \label{eq:22}
\end{eqnarray}
where
\begin{multline}\label{eq:23}
    \rho_{eff}=m^{2}_{eff}H\psi+2\kappa^2f_T(\rho+p)-\kappa^2f \\ -\frac{{m^2}{\psi^2}}{2}-6\lambda\psi^2,
\end{multline}
\begin{multline}\label{eq:24}
    p_{eff}=\frac{m^{2}_{eff}}{3}(\dot{\psi}+\psi^2-4H\psi)+\kappa^2f+4\kappa^2\dot{f_Q}\psi  \\ +\frac{m^2 \psi^2}{2}+6\lambda\psi^2.
\end{multline}
The Universe is filled with dust and radiation fluids,
\begin{equation}\label{eq:25}
    \rho = \rho_m + \rho_r, \ \ \ \ \ p_r=\frac{1}{3}\rho_r,\ \ \ \ p_m=0.
\end{equation}
The density parameters for matter, radiation, and DE components are respectively expressed as,
\begin{equation}\label{eq:26}
\Omega_m=\frac{8\pi G \rho_m}{3 H^2},~~~\Omega_r=\frac{8\pi G \rho_r}{3 H^2},~~~\Omega_{f}=\frac{8\pi G \rho_{eff}}{3 H^2}
\end{equation}
To note, when the parameters approach their limiting values, the gravitational action simplifies the conventional Hilbert-Einstein form. In particular, when $f=0$, $\psi=0$ and $\lambda=\kappa^2$ the effective energy density $(\rho_{eff})$ and effective pressure $(p_{eff})$ become zero and hence the field equations reduce to that of GR. For further analysis, we consider $8\pi G=1$. \\
Eq. \eqref{eq:21} shows that the expansion of the Universe is directly proportional to the Lagrange multiplier $(\lambda)$ and density parameters. So, to analyse the effect on the expansion of the Universe by using $f(Q, T)$ function, we consider $\lambda=\frac{1}{2}$.\\
The following set of dimensionless variables are considered,
\begin{equation}\label{eq:27}
x=\frac{\Psi}{H},\ \ \ y=\frac{Q_0}{Q},\ \ \,z=\frac{\rho_m}{3H^2},\\\,w=\frac{\rho_r}{3H^2}.\ \ \ 
\end{equation}
 Since we need a form of $f(Q,T)$ to frame the cosmological model, here we consider $ f(Q,T)= Q ~e^{\alpha \frac{Q_0}{Q}} + \beta~T$ \cite{Xu_2020_80}. Now, the first Friedmann Eq. (\ref{eq:21}) can be expressed in terms of dynamical variables as,
\begin{multline}\label{eq:28}
    1 = \Omega_m+\Omega_r + \left(\frac{m^2+6(1-\alpha y)e^{\alpha y}+6}{3} \right)x \\ +\beta (z+\frac{4}{3} w) - x^2e^{\alpha y} - \frac{\beta}{2} z - \left(\frac{m^2}{6}+1\right) x^2,
\end{multline}
Using Eq. (\ref{eq:28}), we can write the density parameters in dimensionless variables as,
\begin{equation}\label{eq:29}
\Omega_m=z,\ \ \ \Omega_r=w,\ \ \,\Omega_f=1- \Omega_m -\Omega_r,\ \ \ 
\end{equation}
The second Friedmann Eq. (\ref{eq:22}) can be expressed as,
\begin{multline}\label{eq:30}
    \zeta= -\frac{2\dot H}{3 H^2} = \frac{1}{\left(1+\frac{m^2+6(1-\alpha~y)e^{\alpha y}+6}{6}+2~e^{\alpha y }(\alpha y)^2\right)} \\ \left(1+\frac{1}{3}Z+\left(\frac{m^2+6(1-\alpha~y)e^{\alpha y}+6}{9}\right)(2-7~x+2~x^2)\right. \\ + x^2~e^{\alpha y}+\frac{\beta}{2}~w+\frac{4}{3}~e^{\alpha y}(\alpha y)^2(2-3x+x^2) \\ \left.+\left(\frac{m^2}{6}+1\right)x^2\right).
\end{multline}
Further the deceleration parameter ($q$) and total EoS parameter ($\omega_{tot}$) can be written in terms of dynamical variables as,
\begin{equation}\label{eq:31}
q=-1-\frac{\dot H}{H^2},\ \ \ w_{tot}=-1-\frac{2\dot H}{3 H^2}.\ \ \ 
\end{equation}
The autonomous dynamical system can be obtained by differentiating the dimensionless variables with respect to $N = \ln(a)$ as,
\begin{align}\label{eq.32}
x'&=x^2-\frac{3}{2} \zeta  (1-x)-3 x+2,\nonumber\\
y'&=-\frac{(2 y) \left(-\frac{3 \zeta }{2}+x^2-3 x+2\right)}{x},\nonumber\\
z'&=-z \left(\frac{3 \beta +3}{2 \beta +1}-3 \zeta \right),\nonumber\\
w'&=-w \left(\frac{4 (3 \beta +3)}{5 \beta +3}-3 \zeta \right).
\end{align}
where prime (') denotes differentiation with respect to the e-fold number $N$. The critical points can be obtained by making the system as, $x' = 0$, $y' = 0$, $z'= 0$, and $w' = 0$. The stability of the critical points can be classified as follows: (i) if all eigenvalues are real and negative, then it is a stable node or attractor, whereas if all are positive, it is an unstable node; (ii) if all eigenvalues are real and at least two have opposite signs, it represents the saddle (unstable) point; (iii) if the eigenvalues are complex, then again it is sub-classified as: (a) all eigenvalues having negative real parts lead to stable focus node, while all eigenvalues have positive real parts leads to unstable focus node, (b) if at least two eigenvalues have real parts with opposite signs that leads to saddle focus. The autonomous system [Eqn. \eqref{eq.32}] yields four critical points such as, $C_1$, $C_2$, $C_3$ and $C_4$. TABLE \ref{tableA1} summarizes the critical points along with its existence condition, TABLE \ref{tableA2} summarizes the value of dynamical and geometrical parameters. Brief description of each critical points given below along with the phase portrait.

\begin{table}[H]
    \centering 

    \begin{tabular}{|c |c |c| c| c| c|} 
    \hline\hline 
    \parbox[c][0.9cm]{1.3cm}{Critical Points
    }& ~~~~$x$~~~~ & ~~~~$y$~~~~ & ~~~~$z$~~~~ & ~~~~$w$~~~~ & Exits for \\ [0.5ex] 
    \hline\hline 
    \parbox[c][1.3cm]{1.3cm}{$C_1$ } & $1$ & $0$ & $ 1$ & $0$ & $\beta = 0 $\\
    \hline
    \parbox[c][1.3cm]{1.3cm}{$C_2$ }  & $1$ & $0$ & $ \frac{2}{3 \beta + 2}$ & $0$  & $\beta \neq -\frac{2}{3}$ \\
    \hline
    \parbox[c][1.3cm]{1.3cm}{$C_3$ }  & $1$ &  $0$ & $0$ & $ \frac{4}{5 \beta + 4}$ & $\beta  \neq -\frac{4}{5}$\\
    \hline
    \parbox[c][1.3cm]{1.3cm}{$C_4$ } & $1$ & $y$ & $0$ & $0$ & $y \in \mathbb{R}$\\
 \hline
    \end{tabular}
    \caption{Critical Points and its existence condition.}
    \label{tableA1}
\end{table}

\begin{table}[H]
    \centering 

    \begin{tabular}{|c |c |c| c| c| c|} 
    \hline\hline 
    \parbox[c][0.9cm]{1.3cm}{Critical Points
    }&  $q$ & $\omega_{tot}$ & ~~~$\Omega_m$~~~ & ~~~$\Omega_r$~~~ & $\Omega_{f}$ \\ [0.5ex] 
    \hline\hline 
    \parbox[c][1.3cm]{1.3cm}{$C_1$ } & $1$ & $\frac{1}{3}$ & $0$ & $1$ & $0$\\
    \hline
    \parbox[c][1.3cm]{1.3cm}{$C_2$ }  & $\frac{\beta + 3}{5 \beta + 3}$ & $\frac{1 - \beta}{5 \beta + 3}$ & $0$ & $0$  & $1$ \\
    \hline
    \parbox[c][1.3cm]{1.3cm}{$C_3$ }  & $\frac{1 - \beta}{4 \beta + 2}$, &  $-\frac{\beta}{2 \beta + 1}$ & $\frac{4}{5 \beta+4}$ & $0$ & $1-\frac{4}{5 \beta+4}$\\
    \hline
    \parbox[c][1.3cm]{1.3cm}{$C_4$ } & $-1$ & $-1$ & $0$ & $0$ & $1$\\
 \hline
    \end{tabular}
    \caption{Value of parameters at different critical points.}
    \label{tableA2}
\end{table}

\begin{figure}
    \centering
    \includegraphics[width=6.8cm]{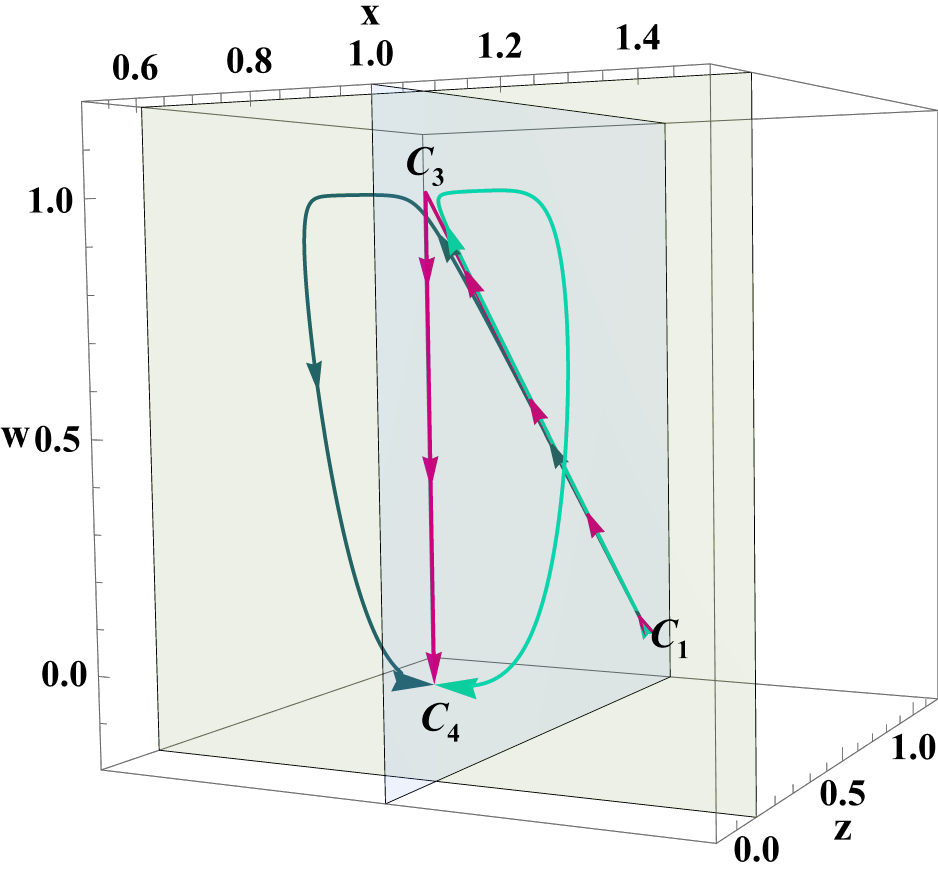}
    \includegraphics[width=6.7cm]{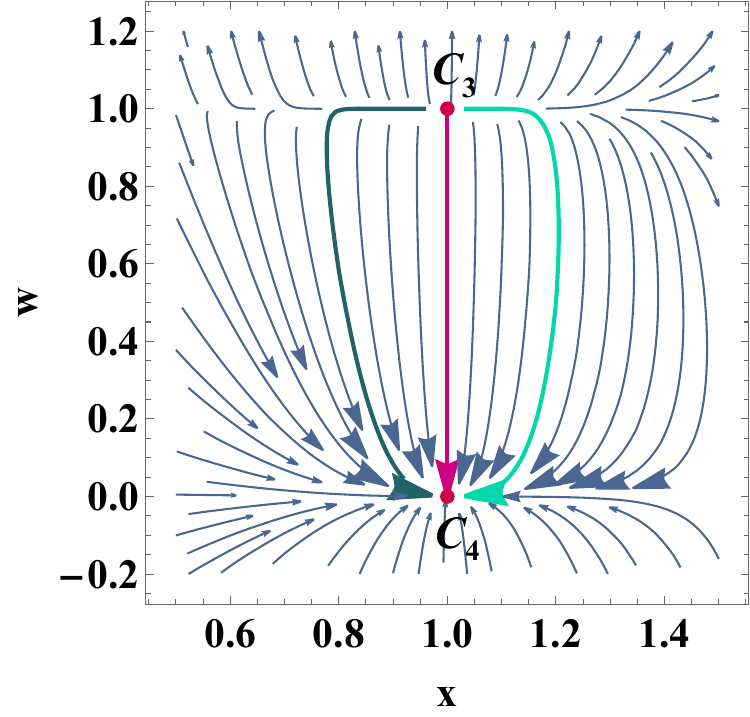}
    \includegraphics[width=6.7cm]{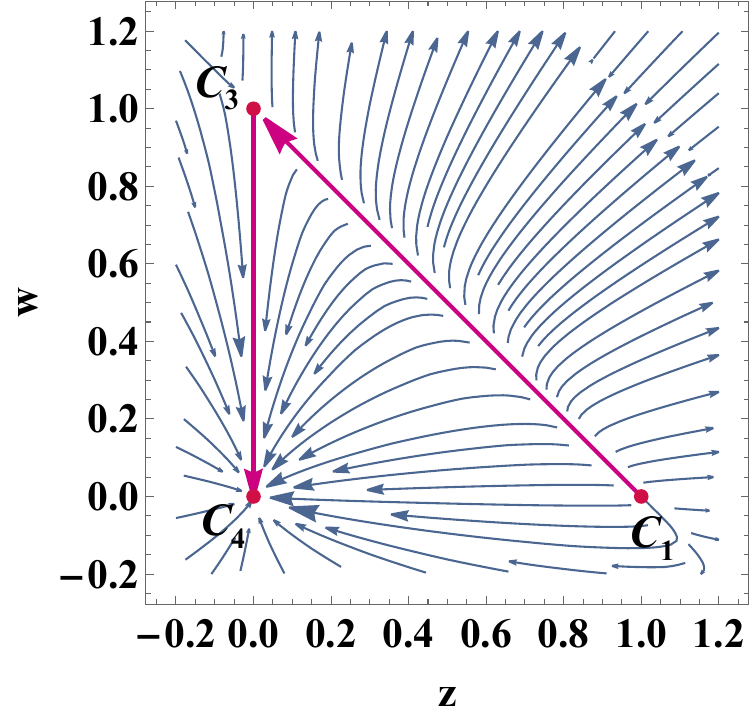}
    \caption{Behaviour of vector in $x$, $z$, and $w$ space and also vectors in xw$-$plane and zw$-$plane.} 
    \label{Fig1}
\end{figure}

\begin{figure}
    \centering
    \includegraphics[width=6.6cm]{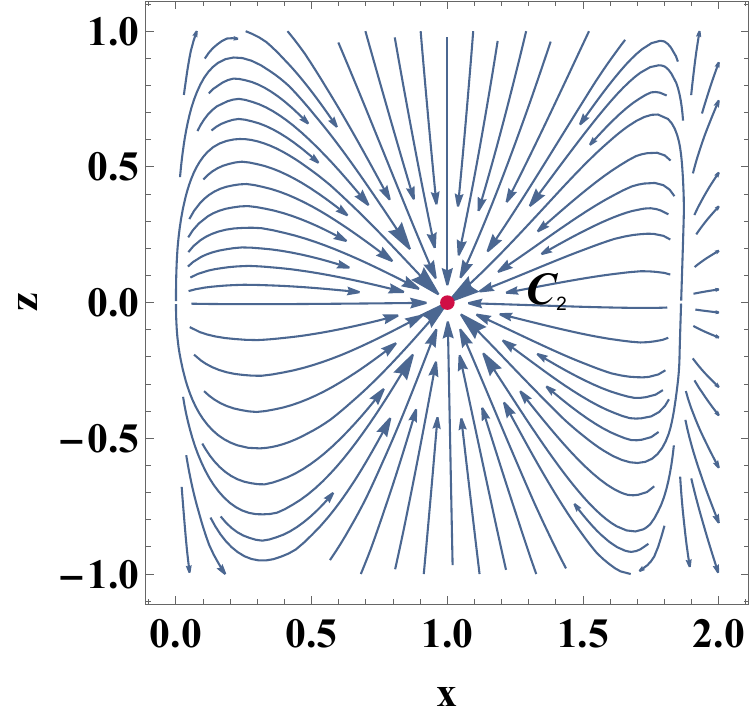}
    \caption{Behaviour critical point $C_2$ in xz-plane.} 
    \label{Fig1a}
\end{figure}

\begin{figure}[H]
    \centering
    \includegraphics[width=6.8cm]{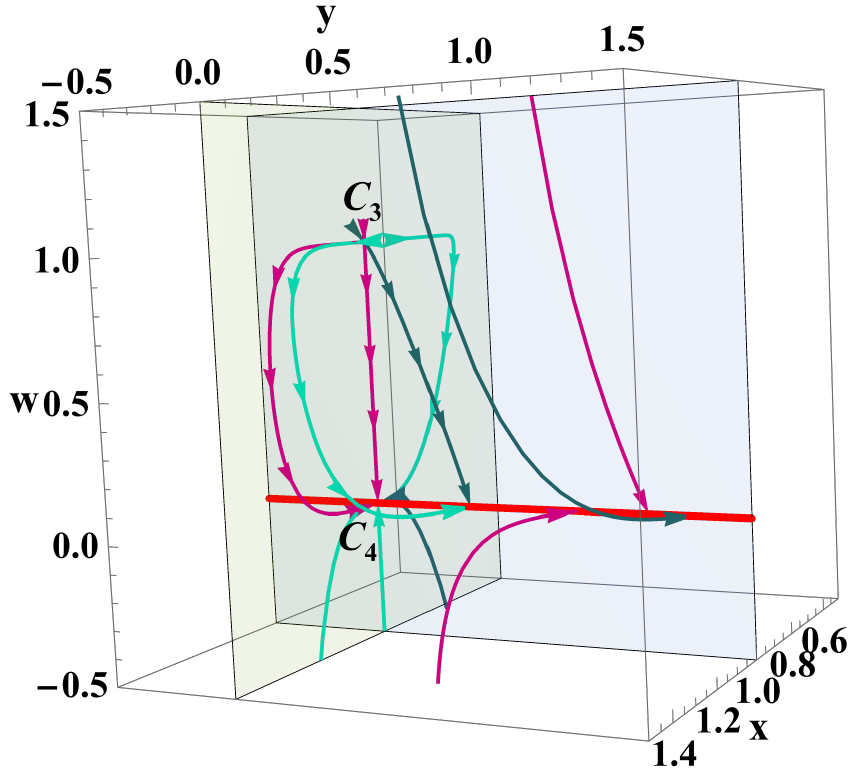}
    \includegraphics[width=6.7cm]{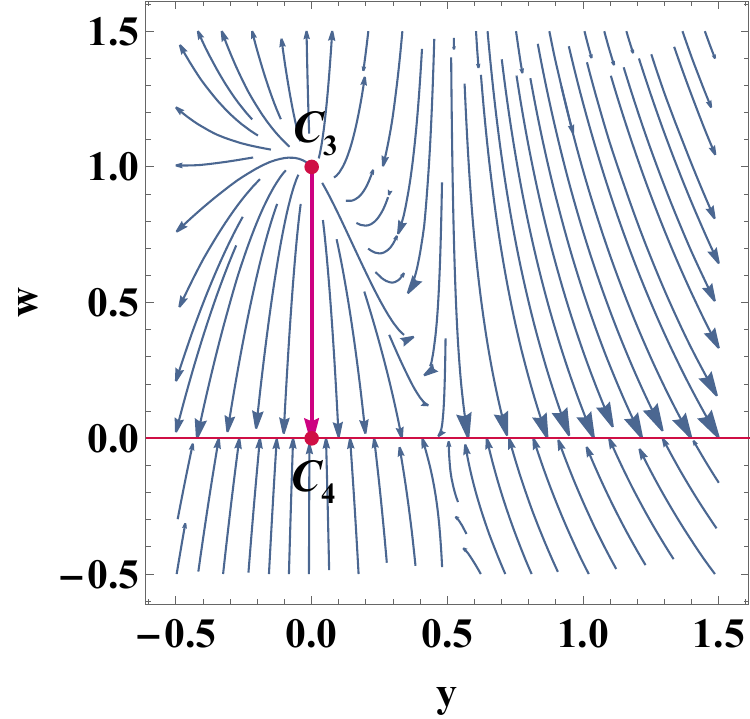}
    \caption{Behaviour of vector in $x$, $y$, and $w$ space and also vectors in wy$-$plane.} 
    \label{Fig2}
\end{figure}

\begin{itemize}
    \item {\bf Critical Point $C_1$($x = 1$, $y = 0 $, $ z = 1$, $w = 0$): } For $\beta=0$, $C_1$ represents the radiation-dominated epoch. The deceleration parameter, $q=1$ and equation of state (EoS) parameter, $\omega_{\text{tot}} = \frac{1}{3}$. The eigenvalues for this critical point are $(8, 4, 1, 1)$, indicating that the system experiences exponential divergence from equilibrium. This can be clearly observed from the phase portrait [FIG. \ref{Fig1}], which represents $C_1$ in the $wz$-plane. It depicts the radiation-dominated phase of the Universe. The non-zero values of $x$ and $w$ at this point are influenced by the non-conservation of the energy balance equation, attributed to the contribution of the Weyl function. The density parameters at this point are obtained as  $\Omega_r = 1, \Omega_m = 0, \Omega_f = 0$ confirms the radiation dominated phase.

    \item {\bf Critical Point $C_2$($x = 1$, $y = 0$, $z = \frac{2}{3 \beta + 2}$, $w = 0$):}
    $C_2$ represents a family of solutions with $q = \frac{\beta + 3}{5 \beta + 3}$ and $w_{tot} = \frac{1 - \beta}{5 \beta + 3}$. The asymptotic solution at $C_2$  describes the transition from the deceleration to the acceleration phase, which depends on the value of the model parameter $\beta$. For different values of $\beta$, the Universe exhibits distinct phases. When $-3\leq \beta \leq -1$,  $-1\leq q\leq0$. The eigenvalues associated with this critical point are,
\[
\left\{\frac{36 (\beta + 1) (3 \beta + 2)}{(5 \beta + 3)^2}, \frac{12 (\beta + 1)}{5 \beta + 3}, \frac{\beta + 3}{5 \beta + 3}, \frac{3 (\beta + 1) (3 \beta + 1)}{(2 \beta + 1)(5 \beta + 3)}\right\}
\]
In the range, $ -1 < \beta < -\frac{2}{3}$, the Universe experiences stable behaviour. In the phase portrait [FIG.\ref{Fig1a}], this stable behavior is exemplified by point  $C_2$, corresponding to the de Sitter Universe. Here, all vectors converge to $C_2$, supporting the stable nature of this point and corresponding density parameter for this point is $\Omega_r=\frac{2}{3 \beta +2}$, $\Omega_m=0$ and $\Omega_f=1-\frac{2}{3 \beta+2}$. 

\item {\bf Critical point $C_3$($x = 1$, $y = 0$, $w = \frac{4}{5 \beta + 4}$, $z = 0$):}
This critical point represents the solution with $q = \frac{1 - \beta}{4 \beta + 2} $ and $ \omega_{tot} = -\frac{\beta}{2 \beta + 1} $. The eigenvalues corresponding for $C_3$ becomes,
\[
\left\{ -\frac{3 (\beta + 1) (3 \beta + 1)}{(2 \beta + 1)(5 \beta + 3)}, -\frac{\beta - 1}{2 (2 \beta + 1)}, \frac{3 (\beta + 1)}{2 \beta + 1}, \frac{3 (\beta + 1)(5 \beta + 4)}{2 (2 \beta + 1)^2} \right\}
\]
For $ \beta = 0 $, $ \omega_{tot} = 0$, which indicates the matter-dominated epoch along with $q=\frac{1}{2}$ indicated decelerating phase of the Universe. The density parameters, $\Omega_m = 1$,  $ \Omega_r = 0$, $\Omega_f=0$. The eigenvalues becomes,$(-1, \frac{1}{2}, 3, 6)$, hence it exhibits saddle behavior during the matter-dominated epoch, The trajectory in the $3-D$ phase portrait in $wxz-$space [FIG. \ref{Fig1}] illustrates the behaviour around the point $C_3$. The system is attracted to $C_3$ along one axis while being repelled along another. This reflects the complex dynamics of the matter-dominated epoch, where the evolution is governed by the attractive and repulsive forces near the critical point.\\

\item {\bf Critical point $C_4$}($x = 1$,$y$, $z = 0$, $w = 0 $). $C_4$ represents the de Sitter phase of the Universe. At this point, $y \in \mathbb{R}$, making the critical point independent of $y$. This phase is characterized by $q = -1$ and $\omega_{\text{tot}} = -1$, the density parameters are $\Omega_m = 0$, $\Omega_r = 0$, and $\Omega_f = 1$. This indicates that the Universe is completely dominated by DE. The eigenvalues for $C_4$ show non-hyperbolic and given as,

\[
\left\{0, -\frac{12 (\beta + 1)}{5 \beta + 3}, -\frac{3 (\beta + 1)}{2 \beta + 1}, -1\right\}
\]

A non-hyperbolic critical point is characterized by the presence of one or more zero eigenvalues, indicating that the linear stability analysis might be inconclusive for some directions. In this case, the eigenvalues include one zero value and three non-zero values, which exhibit negative behavior under the conditions $\beta < -1$ or $\beta > -\frac{1}{2}$. The presence of a zero eigenvalue means that the stability of this point cannot be determined solely from the linearized system and requires further analysis. However, by applying the rule that if the number of vanishing eigenvalues equals the dimension of the center manifold and the remaining eigenvalues indicate normal hyperbolicity, the corresponding critical point is stable \cite{Coley_1999,aulbach_1984_1058,Kadam_2022_82}. In the case of $C_4$, since $y$ is independent and there is only one vanishing eigenvalue, this critical point is considered normally hyperbolic, which suggests the stability of $C_4$. This behaviour indicates that the Universe can remain in this de Sitter phase without diverging away due to small perturbations. In FIG. \ref{Fig2}, the behavior of this point is illustrated in the \( xyw \)-space when  $x = 1$ and  $w = 0$. Here, $y$ demonstrates its free nature, as it can take any value, and all trajectories converge along the line where $x = 1$ and  $w = 0$. This convergence supports the stability of $C_4$, as it shows that despite the freedom in the $y$-direction, the system remains stable. This behavior is further observed in the $wy$-plane, where the trajectories also converge.

\end{itemize}


\begin{figure}[H]
    \centering
    \includegraphics[width=9cm]{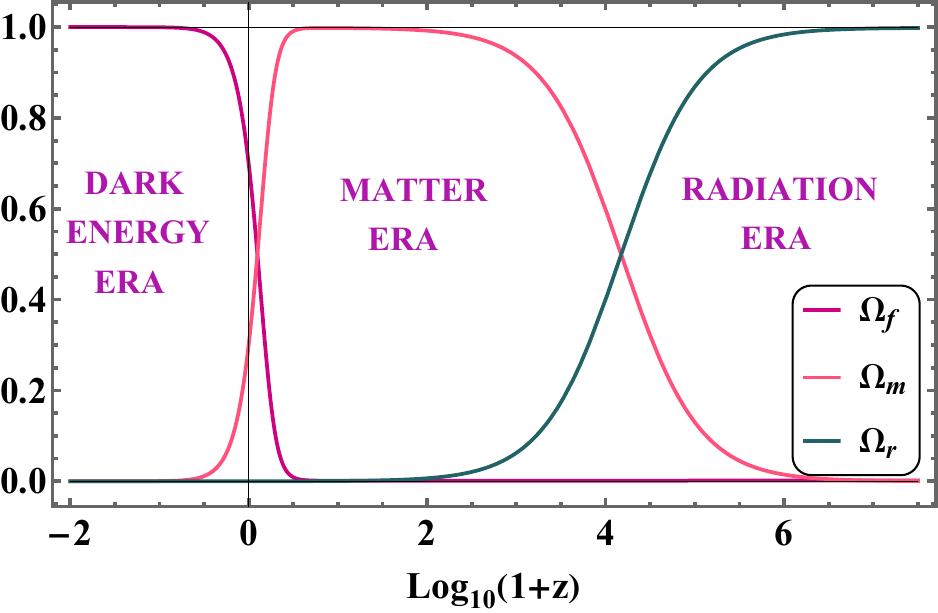}
    \caption{Evolution plots for density parameters; $\alpha=2$, $\beta=-10^{-3}$, $m=2.5$. The initial conditions:$x=10^0$, $y=10^{-5}$, $z=0.32$ and $w=2.47\times10^{-5}$.} 
    \label{Fig3}
\end{figure}

\begin{figure}[H]
    \centering
    \includegraphics[width=9cm]{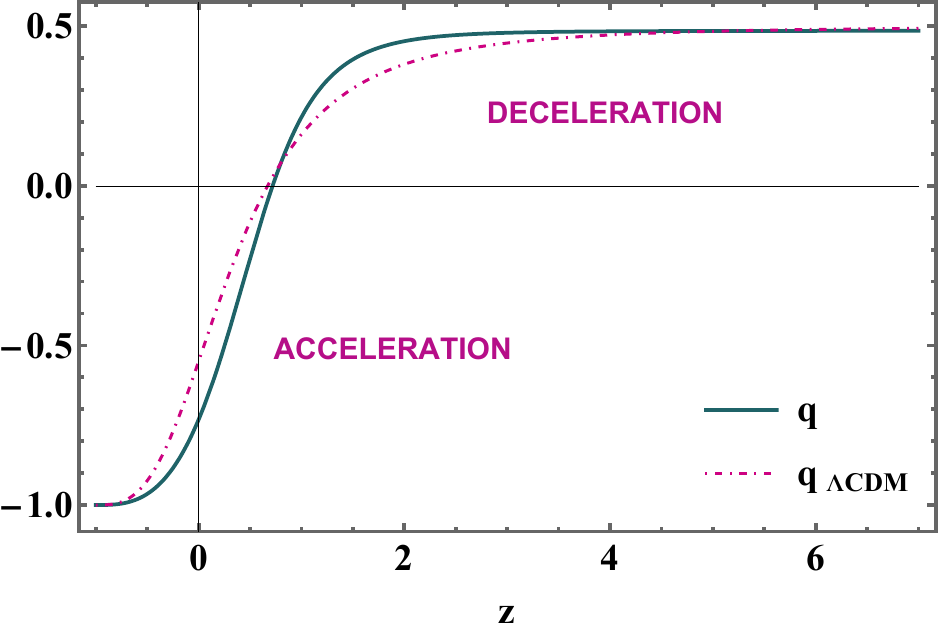}
    \caption{Evolution plots of deceleration parameters} 
    \label{Fig4}
\end{figure}

\begin{figure}[H]
    \centering
    \includegraphics[width=9cm]{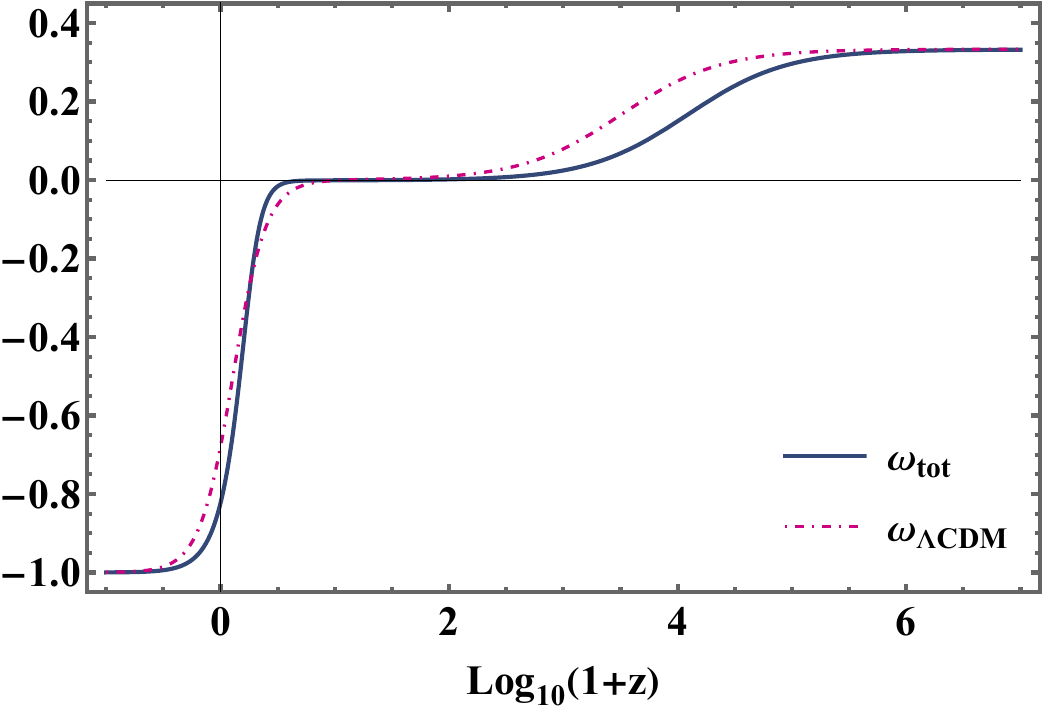}
    \caption{Evolution plots of EoS parameters.} 
    \label{Fig5}
\end{figure}

As the density and geometrical parameters are already expressed in terms of the dimensionless variables. In FIG. \ref{Fig3}, we have illustrated the evolutionary behaviour of density parameters with the initial conditions derived from the dimensionless variables \cite{Bertone_2004_405}. For the model parameter $\alpha$ and the Weyl mass component $m$, we apply the condition $c_1\in \mathbb{Z}\land m\neq 0\land y\neq 0 $ and $ \alpha =\frac{1+2 W_{c_1}\left(\frac{m}{12 \sqrt{e}}\right)}{2 y}$. The parameter $\beta$ is chosen from the range $-5 < \beta < 5$, valid for all critical points, along with $m=2.5$, $\alpha=2$, and $\beta=-10^{-3}$. It has been observed that during the deceleration phase, radiation initially dominates, followed by matter. Eventually, $\Omega_f$ becomes dominant in the de-Sitter phase, where the Weyl-type $f(Q,T)$ function governs the dynamics. The behaviour of deceleration parameter along with the $\Lambda$CDM model has been shown in FIG. \ref{Fig4}. Both the curves transient behaviour with $q_0=-0.716$ and the transition occurs at $z_{tr}=0.714$. At late times both the curves merge and approach to $q=-1$ at late times. The total equation of state parameter is depicted in FIG. \ref{Fig5}, beginning at $\frac{1}{3}$ during the radiation-dominated era, shifting to $w_{tot} = 0$ in the matter-dominated era, and approaching $-1$ at later times, with a present value of $w_{\text{tot}}=-0.835$.

\begin{figure}[H]
    \centering
    \includegraphics[width=9.6cm]{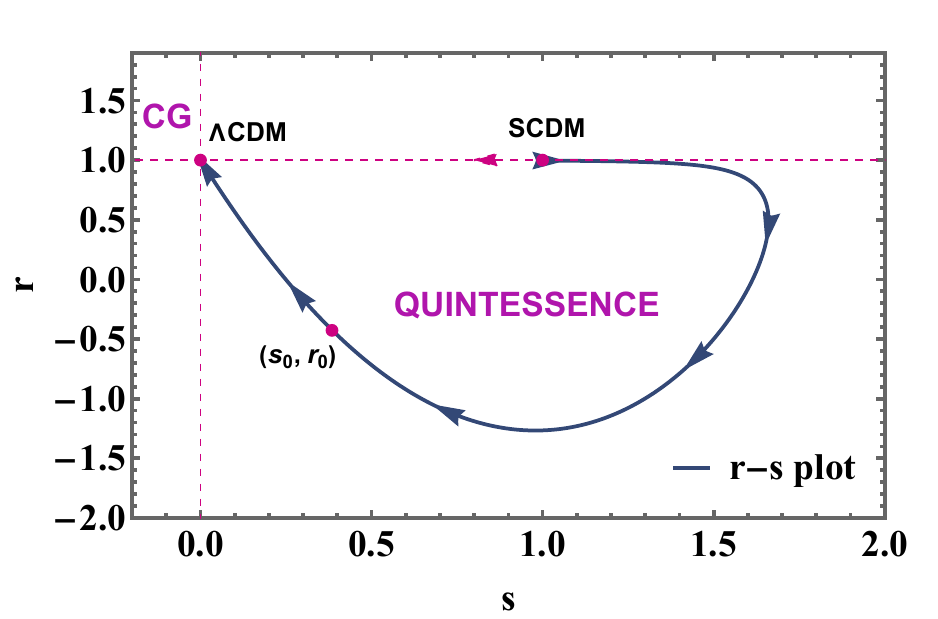}
    \caption{Behaviour of $r-s$ parameter for dynamical system} 
    \label{Fig6}
\end{figure}

To distinguish different DE models, $r-s$ diagnostic is an important tool\cite{Sahni_2003_77}, known as the state finder pair, where $r$ and $s$ respectively be the jerk and snap parameter. These parameters can be obtained as, $r = \frac{\dddot{a}}{aH^3}$ and   $s = \frac{r - 1}{3(q - \frac{1}{2})}$. The graphical representation of $r-s$ space has been given in FIG. \ref{Fig6} and from the trajectory one can see that it includes the $\Lambda$CDM, SCDM, Quintessence and  Chaplygin Gas (CG) model. The $\Lambda$CDM model is marked by a fixed point at $(0, 1)$, which serves as a reference for comparing other models\cite{Alam_2003}. This fixed point reflects the unchanged nature of $\Lambda$ CDM model, with its constant DE component described by the cosmological constant $\Lambda$. In contrast, the SCDM model, which does not include a cosmological constant, follows a different trajectory that diverges from the $\Lambda$CDM point. Further the plot illustrates the paths of SCDM to Quintessence models, characterized by a dynamic scalar field as the DE component where $(s_0, r_0)$ is the present value of $(s,r)$ which moves toward from the $\Lambda$CDM point.

\section{Observational aspects of the  model}\label{Sec:4}

In the previous section, we have performed the dynamical system analysis to understand the cosmic evolution of the Universe for the Weyl type $f(Q,T)$ gravity. Using the dimensionless variables the geometrical and dynamical parameters are presented and analysed for the exponential form of $f(Q,T)$. Here, we shall focus on the use of cosmological datasets for analysing these parameters, so that the comparison can be made in both the approaches. We can see that, the non-metricity $( Q )$ depends on $( \psi )$, which is governed by Eqn. \eqref{eq:14}, and the Friedmann equations \eqref{eq:17} and \eqref{eq:18}, derived from varying the action with respect to the FLRW metric. By incorporating $f(Q, T) = Q ~e^{\alpha \frac{Q_0}{Q}} + \beta~T $, we obtain the corresponding equations in terms of the redshift form Eq. \eqref{eq.33} to Eq. \eqref{eq.36}. These equations will be central to our observational analysis, where we utilize the Cosmic Chronometers ($CC$) and $Pantheon^+$ dataset to constrain the model parameters and validate our theoretical findings. Through this approach, we aim to explore the compatibility of the Weyl-type \( f(Q, T) \) gravity with the late-time acceleration of the universe and its overall consistency with current cosmological observations. We redefine the independent variable by introducing the redshift \( z \), instead of the time variable \( t \). The redshift is defined by the relation:

\[
1 + z = \frac{1}{a}
\]

where the scale factor \( a \) has been normalized to unity at the present time, \( a(0) = 1 \). As a result, the time derivative becomes:

\[
\frac{d}{dt} = \frac{dz}{dt} \frac{d}{dz} = -(1 + z)H(z) \frac{d}{dz}
\]

This expresses the time operator in terms of the redshift \( z \) and the Hubble parameter \( H(z) \).

\begin{widetext}
\begin{equation}\label{eq.33}
    -(z+1) H \frac{\text{d$\psi $}}{\text{dz}}=-(z+1) H\frac{\text{dH}}{\text{dz}}+2 H^2-3 H \psi +\psi ^2
\end{equation}
\begin{equation}\label{eq.34}
    (z+1) H\frac{\text{d$\lambda $}}{\text{dz}}=\frac{1}{6} \psi  \left(6 e^{\alpha\frac{  \psi _0^2}{\psi ^2}} \left(1-\alpha\frac{  \psi _0^2}{\psi ^2}\right)+12 \lambda +m^2\right)
\end{equation}
\begin{multline}\label{eq.35}
    (z+1) H\frac{\text{dH}}{\text{dz}}=\frac{1}{12 (\beta +2) \lambda  \psi ^4}\left(\psi ^4\left(12 \lambda  \left((2 \beta +1) \psi ^2+3 (\beta +1) H^2-(5 \beta +4) H \psi +(\beta +2)(z+1) H \frac{\text{d$\psi $}}{\text{dz}}\right)\right.\right.\\\left.m^2 (\psi  (2 \beta  \psi -5 \beta  H-4 H+\psi )+(\beta +2) (z+1) H \frac{\text{d$\psi $}}{\text{dz}})++6 e^{\frac{\alpha  \psi _0^2}{\psi ^2}} (\psi  (2 \beta  \psi -5 \beta  H-4 H+\psi )+(\beta +2) (z+1) H \frac{\text{d$\psi $}}{\text{dz}})\right) \\ \left.-6 \alpha  \psi _0^2 e^{\frac{\alpha  \psi _0^2}{\psi ^2}} \left(-2 \alpha  (\beta +2) \psi _0^2 (z+1) H \frac{\text{d$\psi $}}{\text{dz}}-\psi ^2 (\psi  ((\beta +2) \psi +(5 \beta +4) H)-(\beta +2) (z+1) H \frac{\text{d$\psi $}}{\text{dz}})\right)\right)
\end{multline}
\begin{equation}\label{eq.36}
    \rho =-\frac{\frac{1}{2} (-3) \psi ^2 e^{\frac{\alpha  \psi _0^2}{\psi ^2}}-3 \lambda  \left(H^2+\psi ^2\right)+\frac{1}{2} H \psi  \left(6 e^{\frac{\alpha  \psi _0^2}{\psi ^2}} \left(1-\frac{\alpha  \psi _0^2}{\psi ^2}\right)+12 \lambda +m^2\right)-\frac{m^2 \psi ^2}{4}}{\frac{\beta }{4}+\frac{1}{2}}
\end{equation}
\end{widetext}

\subsection{MCMC Analysis}
We shall solve the system of equations numerically. To estimate the parameters $H_0$, $\alpha$, $\beta$, $m$, and $\psi_0$, we employed Markov Chain Monte Carlo (MCMC) analysis \cite{Foreman-Mackey_2013_125}. The initial value for $\lambda_0 = 0.5$ has been taken as in the theoretical analysis. The minimum chi-squared ($\chi^2_{\text{min}}$) value is a crucial metric in cosmology, indicating the best-fit parameters that minimize the difference between observational data and model predictions. The chi-squared statistic for 32 points for $CC$ dataset \cite{Moresco_2022_25} from redshift $0.07$ to $1.965$ is given as,
\begin{equation}\label{eq.37}
\chi^2_{CC} = \sum_{i=1}^{32} \left( \frac{H_{\text{obs},i} - H_{\text{model},i}}{\sigma_i} \right)^2,
\end{equation}
where $H_{\text{obs},i}$ and $H_{\text{model},i}$ respectively be the observed and model-predicted Hubble parameters at the $i$-th redshift point, $\sigma_i$ is the observational uncertainty. The log-likelihood function, essential in MCMC analysis and can be expressed as,

\begin{equation}\label{eq.38}
\log \mathcal{L} = -\frac{1}{2} \chi^2
\end{equation}
The posterior probability combines the prior and the likelihood as,

\begin{equation}\label{eq.39}
\log \mathcal{P}(\theta | \text{data}) = \log \mathcal{L}(\text{data} | \theta) + \log \pi(\theta),
\end{equation}
where $\theta$ represents the model parameters. A lower $\chi^2_{\text{min}}$ value suggests a better fit and helps in validating the model and quantify the uncertainties in parameters.
Supernova data from Type Ia supernovae has been  crucial to determine the cosmological parameters. The system is numerically integrated over a range of redshifts to obtain $H(z)$. The $Pantheon^+$ dataset comprises 1701 Type Ia supernova measurements form redshift $0.00122$ to $2.2613$. The chi-square statistic for this dataset is,

\begin{equation}\label{eq.40}
    \chi^2_{{Pantheon^+}}(\theta) = \left( \vec{\mu}(\theta) - \vec{\mu}_{\text{obs}} \right)^T C_{{Pantheon^+}}^{-1} \left( \vec{\mu}(\theta) - \vec{\mu}_{\text{obs}} \right),
\end{equation}
where
\begin{itemize}
    \item $\vec{\mu}(\theta)$ is the theoretical distance moduli,
    \[
    \mu_{\text{model}}(z) = 5 \log_{10}(d_L(z)) + 25
    \]
    \item $d_L(z)$ is the luminosity distance,
    \[
    d_L(z) = (1+z) \int_0^z \frac{c \, dz'}{H(z')}
    \]
    \item $\vec{\mu}_{\text{obs}}$ is the observed distance moduli,
    \item $C_{Pantheon^+}$ is the covariance matrix of the observed data.
\end{itemize}

Also, $(\sigma_i)$ represents the uncertainties in the observed distance moduli, the parameters \(H_0\), \(\alpha\), \(\beta\), \(m\), and \(\psi_0\) are adjusted to minimize the chi-squared value using MCMC methods to obtain the best fit to the data. The prior is  $H_0 \in(60, 80)$ \cite{Rebecca_2022} and the model parameters $\alpha \in(-5,5)$, $\beta \in(-5,5)$. For the Weyl mass vector, we take the range $(0, 5)$  \cite{Xu_2020_80}, and for $(\psi_0)$, we used the same range as for \(H_0\).

\begin{widetext}

\begin{figure}[H]
\centering
\includegraphics[width=8.5cm,height=9cm]{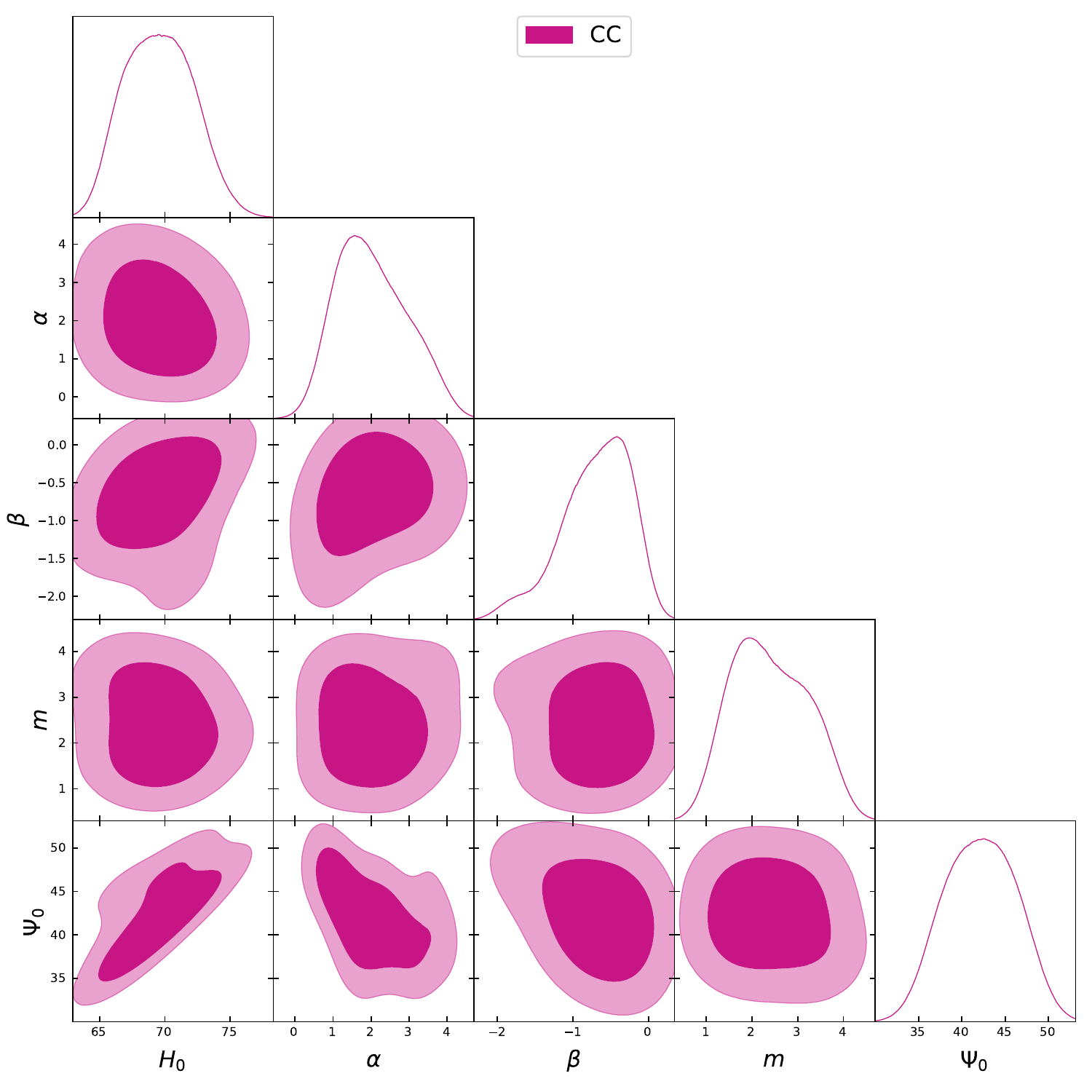}
\includegraphics[width=8.5cm,height=9cm]{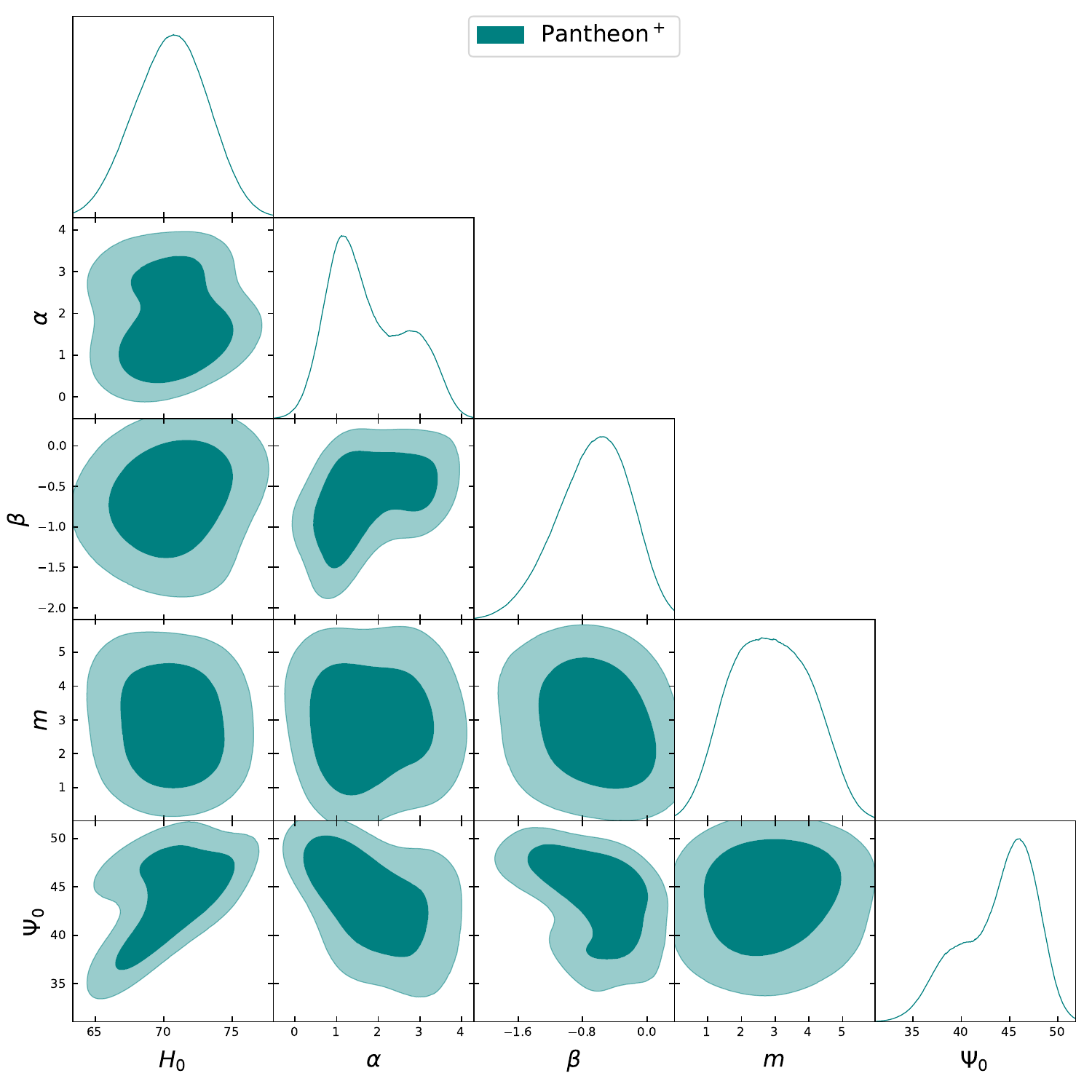}
\caption{Contour plot obtained from $CC$  {\bf (Left Panel)} and $Pantheon^{+}$ datasets {\bf (Right Panel)}.}
\label{Fig7}
\end{figure}

\begin{figure}[H]
\centering
\includegraphics[width=8.5cm,height=6cm]{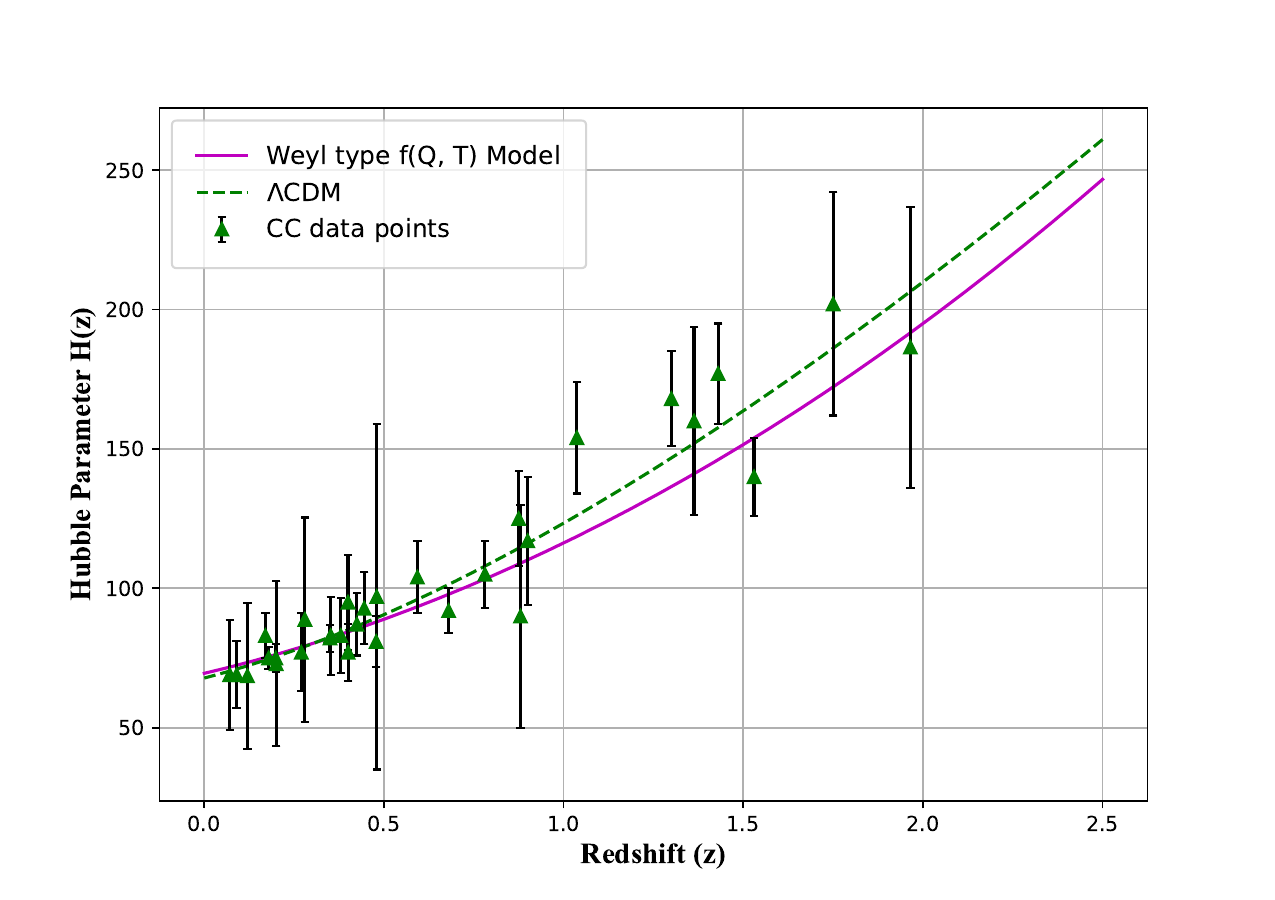}
\includegraphics[width=8.5cm,height=6cm]{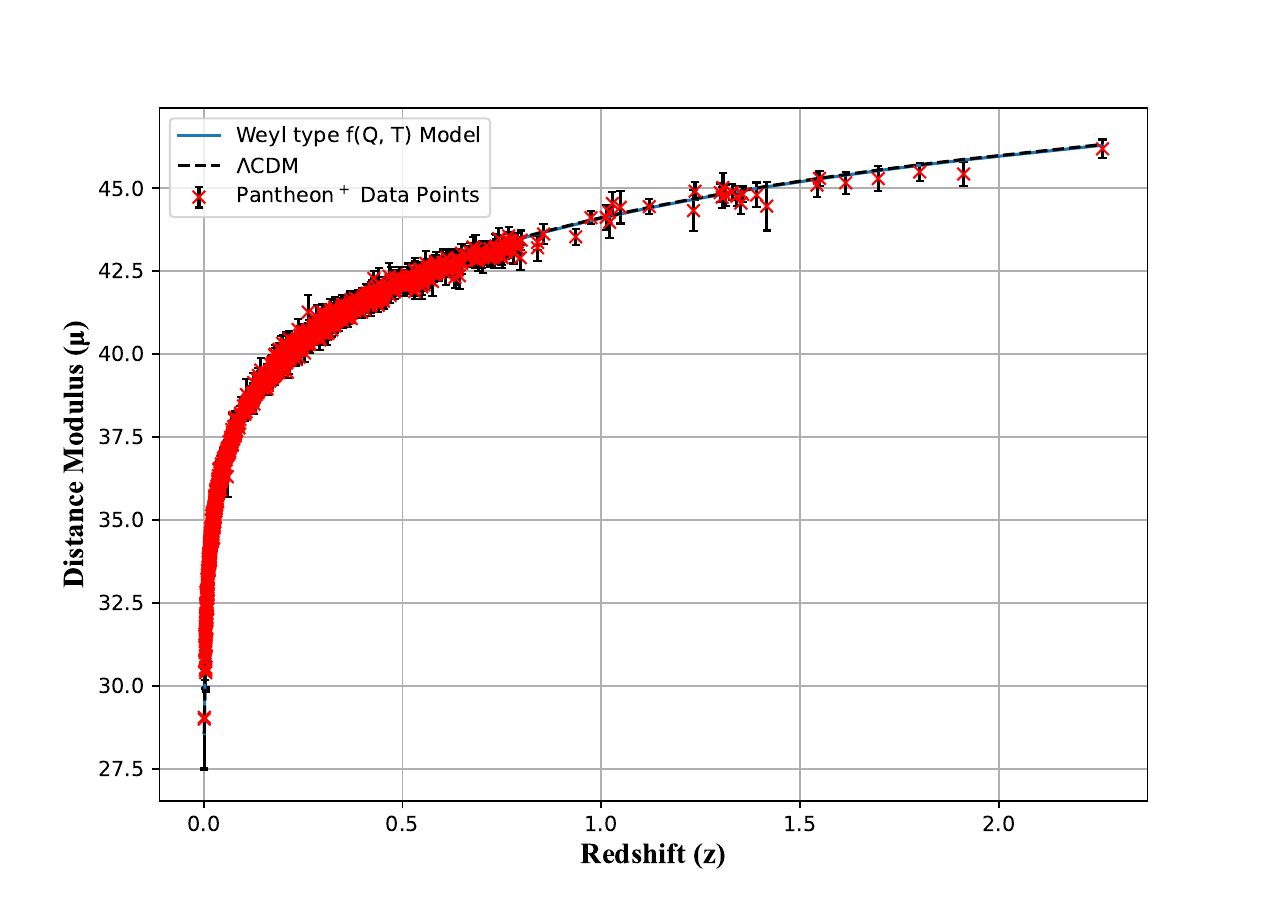}
\caption{Evolution of Hubble parameter and distance modulus parameter for the model, $\Lambda$CDM and $CC$  {\bf (Left Panel)} and $Pantheon^{+}$ datasets {\bf (Right Panel)}}
\label{Fig8}
\end{figure}
\end{widetext}

\subsection{Results from Observational Data}

The corner plot of the posterior distributions, shown in FIG. \ref{Fig7}, demonstrates that these parameters are well-constrained with relatively narrow uncertainties. Notably, the parameter values derived from our theoretical analysis fall within the uncertainty ranges provided by the observational data. The best-fit parameters, as presented in Table \ref{Table-I}, yielded minimum chi-squared values of \(\chi^2_{\text{min}} = 16.92\) for the $CC$ data and \(\chi^2_{\text{min}} = 1119.19\) for the $Pantheon^+$ data, indicating a strong concordance between the model and observational evidence. 

\begin{table*}
\renewcommand\arraystretch{1.5}
\centering 
\begin{tabular}{|c|c|c|c|c|c|c|c|c|c|} 
\hline\hline 
~~~Datasets~~~& ~~~$H_0$  ~~~& ~~~~$\alpha$~~~~ & ~~~$\beta$~~~ &$m$&$\psi_0$& \parbox[c][1.5cm]{2cm}{Age of the Universe}&$\chi^2_{min}$&$\chi^2_{min(\Lambda CDM)}$&$\triangle\chi^2_{min}$
     \\ [0.5ex] 
\hline\hline
\textit{$CC$} dataset & 69.5 $\pm$ 2.6 & 2.02 $^{+0.79}_{-1.1}$ &  $-0.69^{+0.74}_{-0.54}$&$2.39^{+0.79}_{-0.99}$& 42.2 $\pm$ 4.0 &$13.81$ Gyr& $16.92$&$14.55$& $2.37$\\
\hline
\textit{$Pantheon^+$} & $70.6\pm2.5$ &  $1.76\pm0.93$ & $-0.66^{+0.71}_{-0.50}$& $2.9\pm1.1$&$44^{+5}_{-3}$&$13.96$ Gyr&$1119.19$& $1093.001$& $26.18$ \\
\hline
\textit{Prior Ranges} & $(60,80)$ &  $(-5,5)$ & $(-5,5)$& $(0,5)$&$(60,80)$&$-$&$-$&$-$&$-$\\
\hline 
\end{tabular}
\caption{Constrained Parameter Values, Universe Age, and $\chi^2_{min}$ for $CC$ and $Pantheon^+$ Datasets and its difference with $\Lambda$CDM.} \label{Table-I}
\end{table*}

The comparison with the standard $\Lambda$CDM model suggests that the $f(Q, T)$ model offers a comparable or potentially superior fit to the data, as demonstrated in FIG. \ref{Fig8} \textbf{(Left Panel)} for the $CC$ dataset, with error bars included. This result highlights the $f(Q, T)$ model's potential as a viable alternative cosmological model. The findings emphasize the effectiveness of MCMC methods in the estimation of cosmological parameters and model comparison. In FIG. \ref{Fig8} \textbf{(Right Panel)}, the distance modulus plot using the Pantheon$^+$ dataset, which contains 1701 data points, is presented. The model curve closely mimics the behavior of the $\Lambda$CDM model.\\

\begin{figure}[H]
    \centering
    \includegraphics[width=9cm]{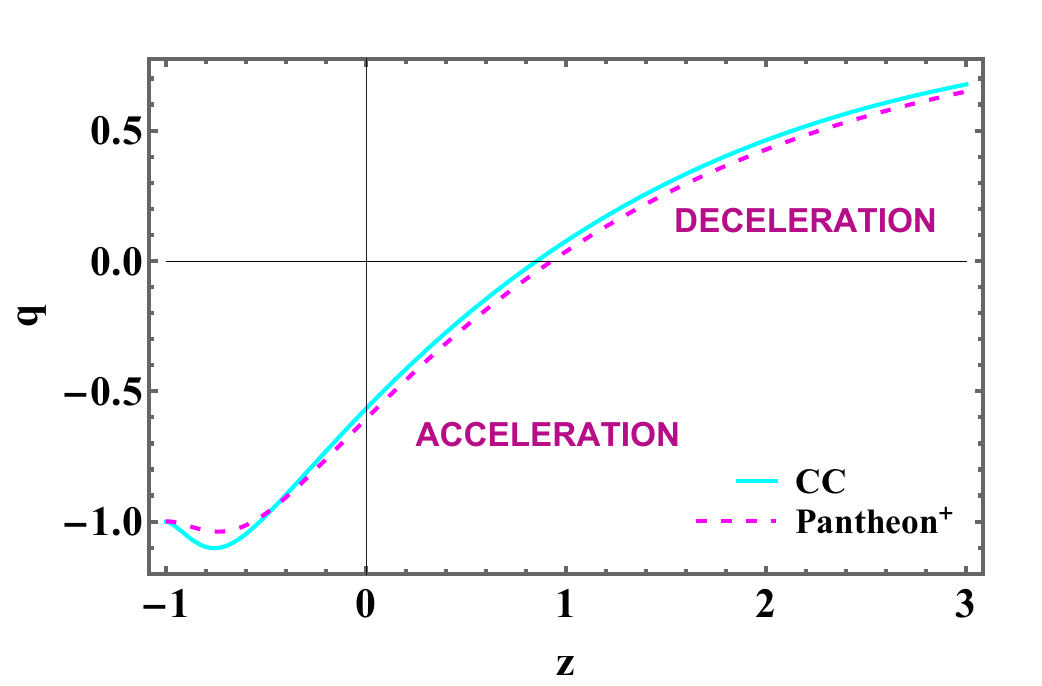}
    \caption{Deceleration parameter in redshift.} 
    \label{Fig11}
\end{figure}

\begin{figure}[H]
    \centering
    \includegraphics[width=9cm]{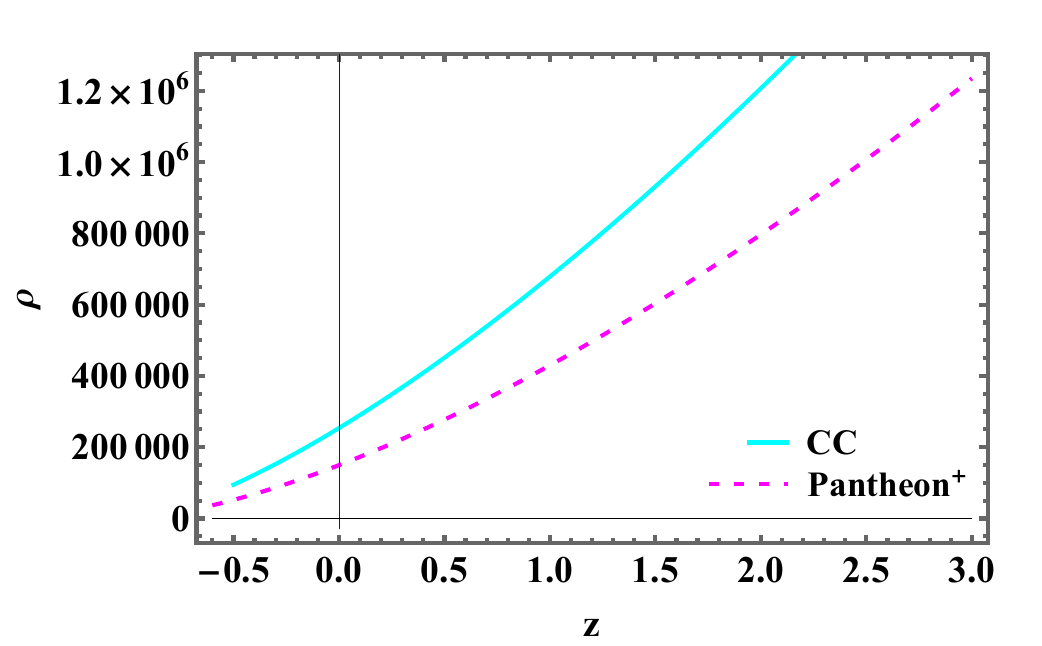}
    \caption{Energy density in redshift.} 
    \label{Fig12}
\end{figure}

\begin{figure}[H]
    \centering
    \includegraphics[width=9cm]{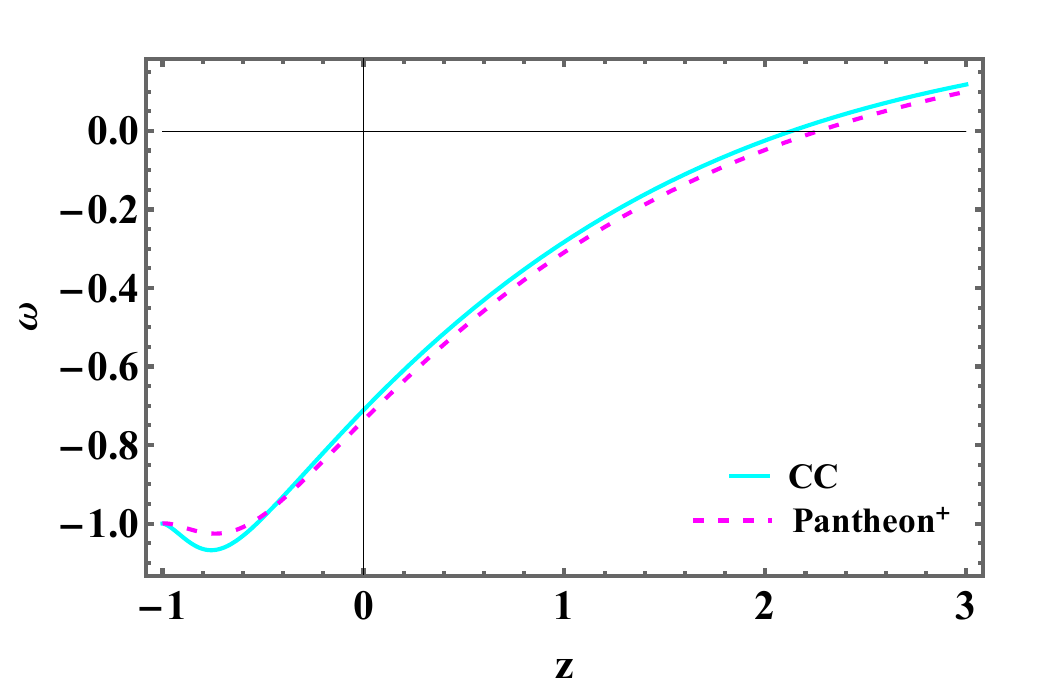}
    \caption{EoS parameter in redshift.} 
    \label{Fig13}
\end{figure}

Using the constrained values of the parameters, the behaviour of deceleration parameter given in FIG. \ref{Fig11}. The curve  indicates the  transitions from decelerating to accelerating phase with the transition respectively at $z_r = 0.852$ and $z_r = 0.934$ for $CC$ and $Pantheon^+$ datasets.  Also, the present value of respectively noted as $q_0=-0.568$ and $q_0=-0.611$ for $CC$ and $Pantheon^+$. At late times both the curves approaches to $-1$. FIG. \ref{Fig12} shows that the energy density remains positive throughout the evolution and reducing from a higher to lower value. The EoS parameter shown in FIG. \ref{Fig13} and the present value observed to be $-0.712$ and $-0.740$ respectively for $CC$ and $Pantheon^+$ dataset. 

\begin{figure}[H]
\centering
\includegraphics[width=8.6cm]{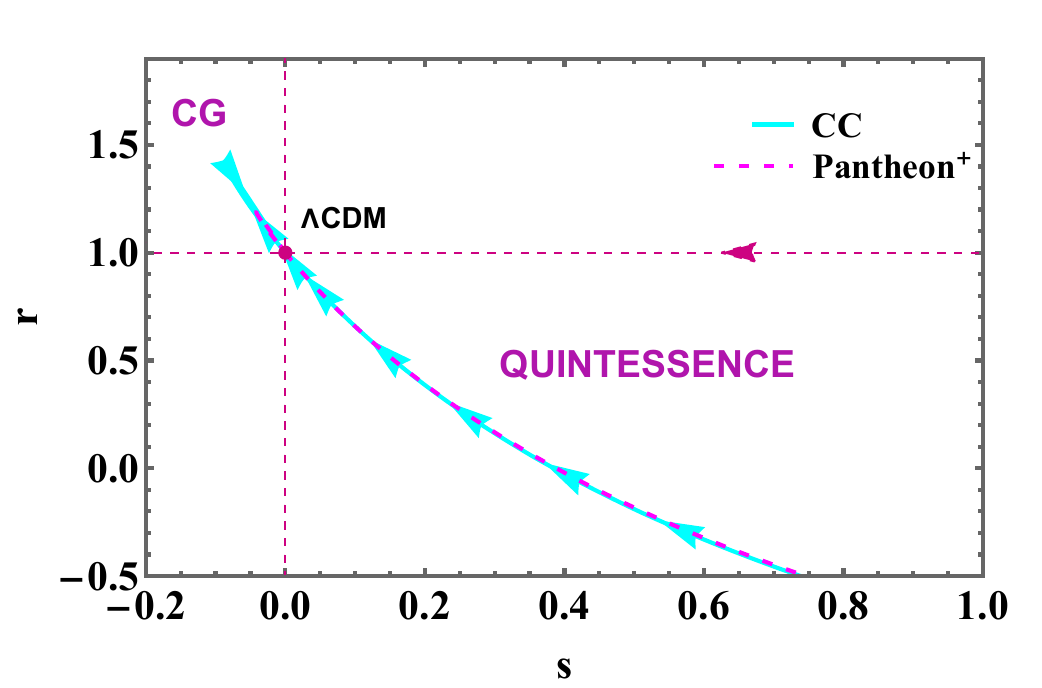}
\includegraphics[width=8.2cm]{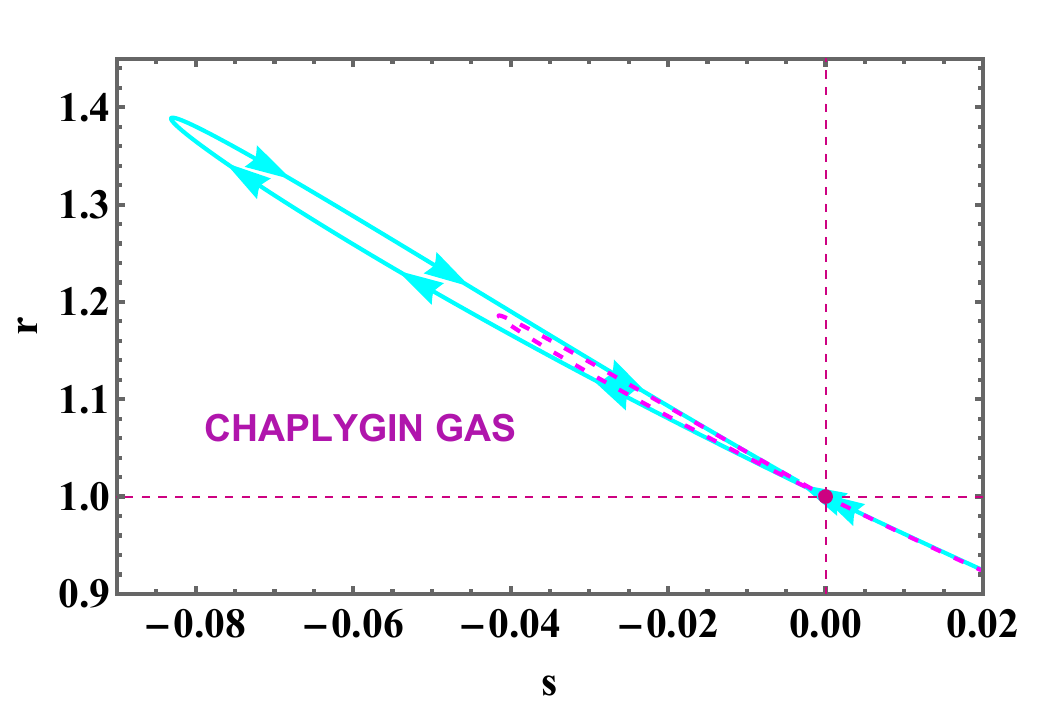}
\caption{Statefinder pair $(r,s)$ in redshift. } 
\label{Fig14}
\end{figure}

FIG. \ref{Fig14} illustrates the behavior of the  $r-s$ pair with the constrained values from  $CC$ and $Pantheon^+$ datasets. In the upper panel, one can observe that the curve traverses from quintessence to $\Lambda$CDM and finally remains at Chaplygin gas. Whereas at the lower panel, we have shown the expanded behaviour  Chaplygin gas. The $\Lambda$ has been confirmed with the value of the pair $(r,s)$ coming close to $(r,s)=(1,0)$ at present. At the same time $r>1, s<0$ suggests the Chaplygin gas state, where the expansion accelerates more rapidly than in standard $\Lambda$CDM model. Conversely, for $r<1, s>0$, it indicates the quintessence phase. While examining the $(r,s)$ pair, we observed a smooth transition from quintessence to $\Lambda$CDM, extending to Chaplygin gas, and converging with $\Lambda$CDM. This behavior signifies the ability of the model to capture late-time cosmic dynamics.
We shall now calculate the age of the Universe considering the constrained values of the parameters obtained through  $CC$ and $Pantheon^+$ datasets [TABLE-\ref{Table-I}]. The age of the Universe can be determined by evaluating the integral \( \int_{0}^{\infty} \frac{{H_0}}{(1+z) H(z)}~dz \), which effectively sums the contributions of the entire expansion history of the Universe. Using the constrained values of the parameters and acknowledging the negligible effect from radiation density, the age of the Universe obtained as $\approx13.81 Gyr$ and $\approx13.96 Gyr$ respectively for $CC$ and $Pantheon^+$ dataset. The average age from various chronometric pairs suggested to be $\approx 13.8 \pm 4\ Gyr$ \cite{Cowan_2002}.

 \begin{figure}
    \centering
    \includegraphics[width=9.1cm]{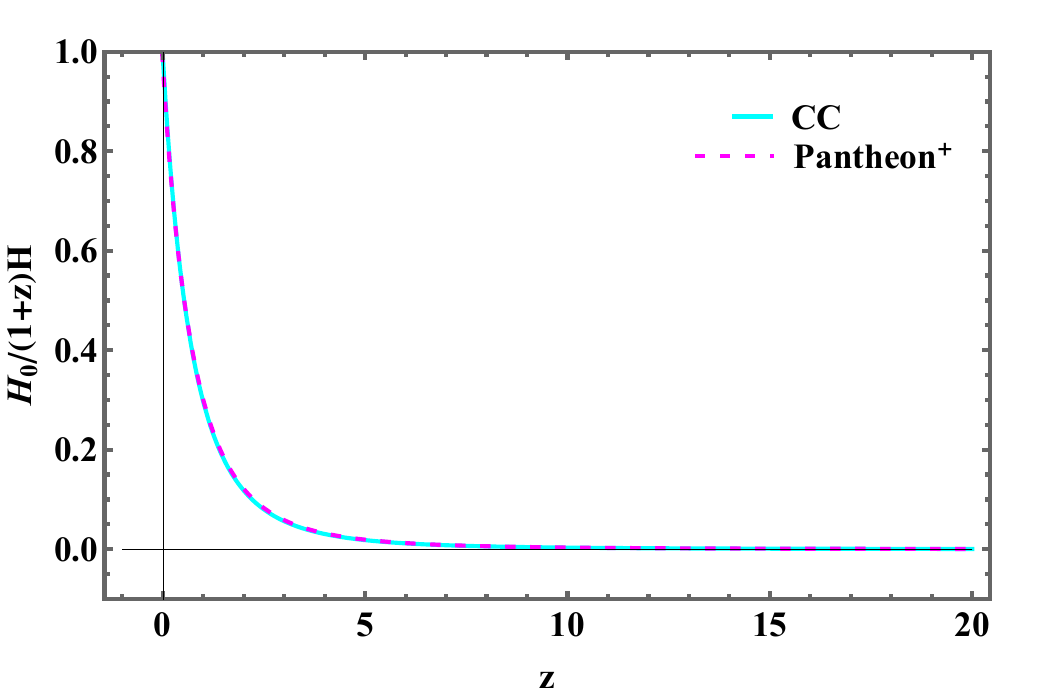}
    \includegraphics[width=9cm]{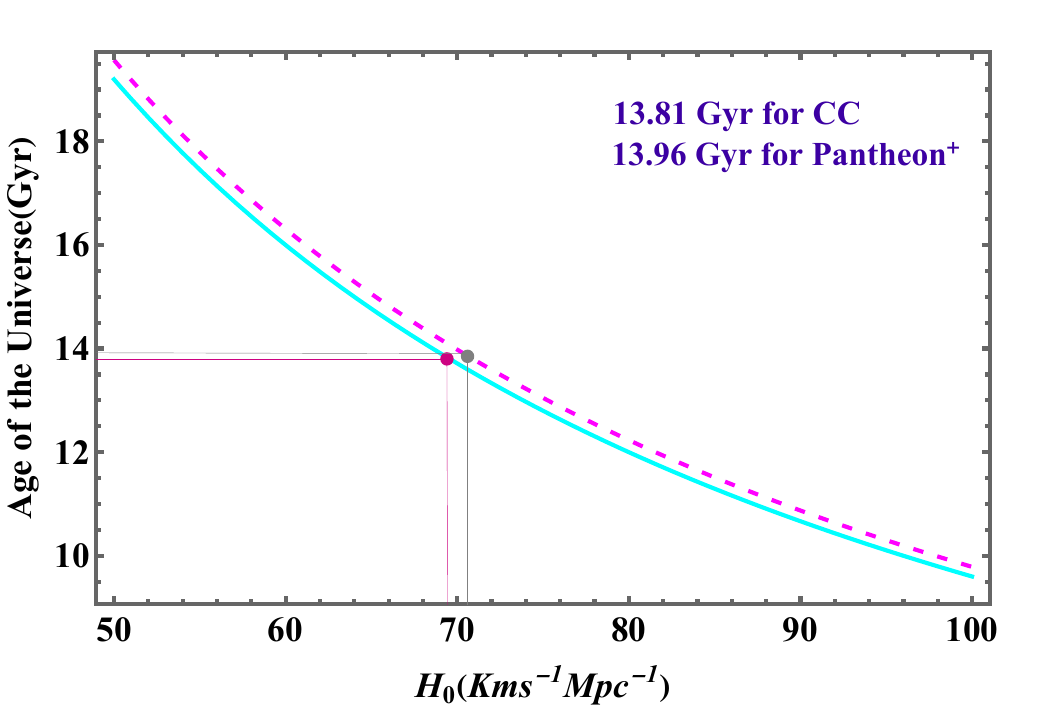}
    \caption{Evolution $H_0/(1+z)H$ vs redshift which shows the conversion to zero \textbf{(Upper Panel)} and \textbf{(Lower Panel)} Shows Age of the Universe with respect to expansion rate. } 
    \label{Fig15}
\end{figure}

\section{Summary and Conclusion}\label{Sec:5} 

\begin{figure}
    \centering
    \includegraphics[width=10cm]{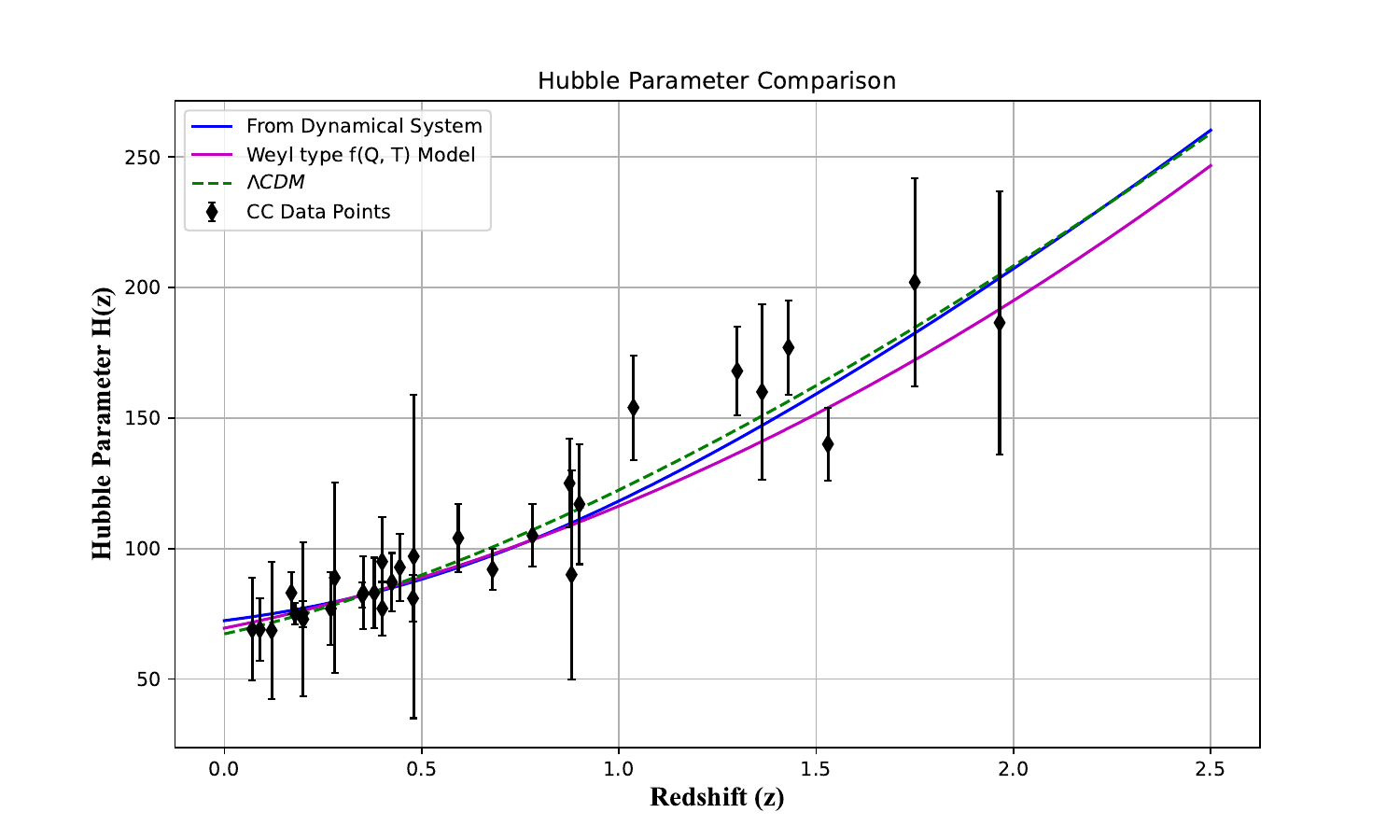}
    \caption{Comparison plot of $H(z)$ between dynamical system, Weyl model and $\Lambda$CDM with observational $32-$ $CC$ datasets.} 
    \label{Fig16}
\end{figure}

\begin{figure}
    \centering
    \includegraphics[width=10cm]{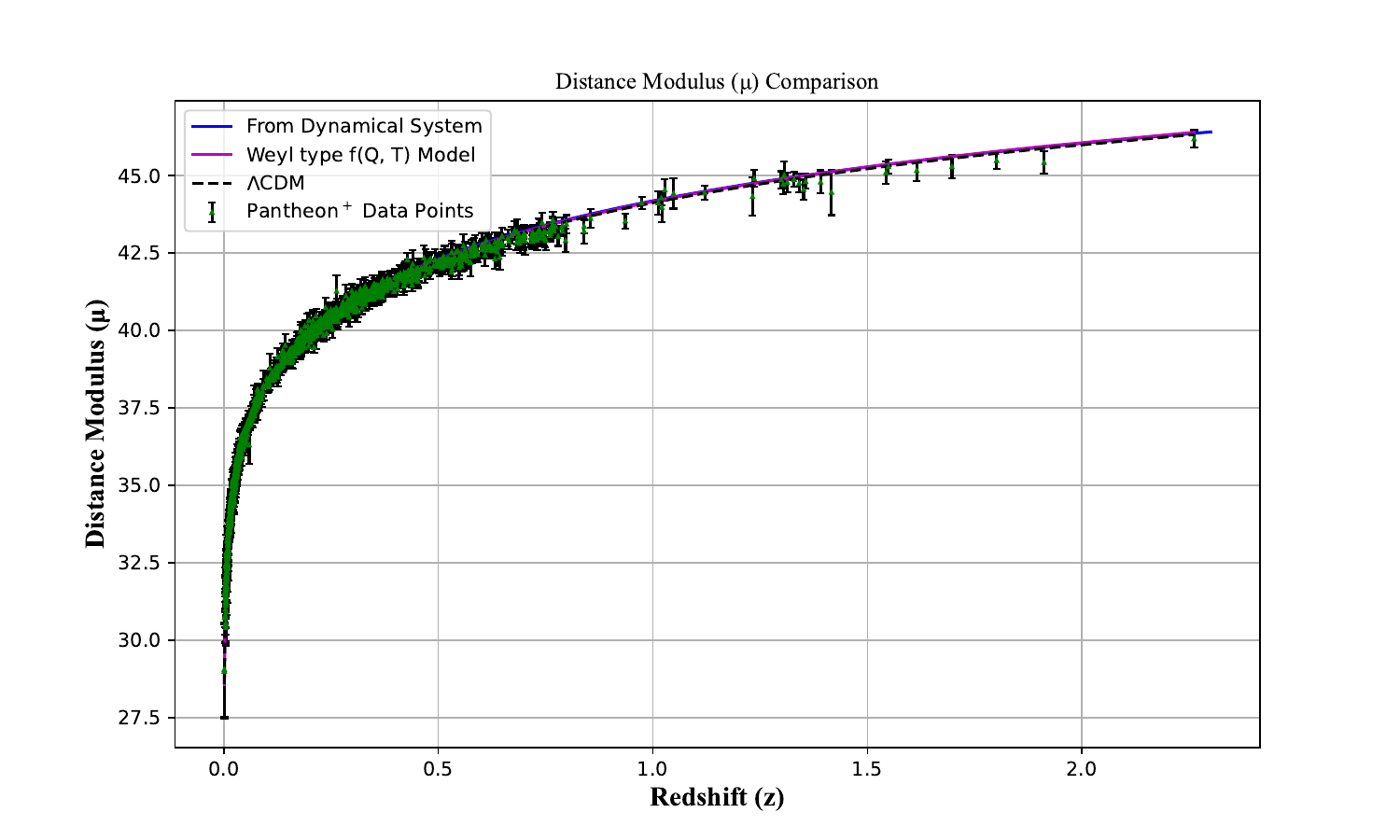}
    \caption{Comparison plot of distance Modulus $(\mu)$ between dynamical system, Weyl model and $\Lambda$CDM with $1701-$ $Pantheon^+$ datasets.} 
    \label{Fig17}
\end{figure}

In Weyl type $f(Q,T)$ gravity, late time accelerating behaviour of the Universe has been presented in the paper with the dynamical system analysis approach and using cosmological datasets. We have focused on the exponential form of the functional$f(Q,T)$. In the first phase, we have adopted the dynamical system approach and define the dimensionless variables to obtain autonomous system and subsequently the critical points. Each critical points obtained are analysed and identified their presence in the evolution eras both from the obtained eigenvalues and the phase portraits. Further the critical points are arranged in such a manner that it shows radiation dominated to matter dominated and de-Sitter phase of the Universe. Further with the initial conditions of the dimensionless variables the deceleration parameter, EoS parameters and density parameters are presented and from its evolutionary behaviour the cosmic acceleration at late times has been confirmed.
In the second phase, we have used the $CC$ and $Pantheon^+$ datasets to constrain the Hubble and model parameters for the exponential form of the functional $f(Q,T)$. After obtaining the constrained values of these parameters, we have analysed the dynamical and geometrical parameters of the model. The deceleration parameter show a clear transition from decelerating to accelerating phase and the EoS parameter reveals the alignment of the model to $\Lambda$CDM at late times. 
Finally we have compared the results obtained through both the approaches. Table$-$\ref{Table-II} summarizes the values of the parameters obtained in both the approaches, which shows in the acceptable range. FIG. \ref{Fig16} provides the plot of $H(z)$ obtained through dynamical system and $CC$ datasets; whereas FIG. \ref{Fig17} for $Pantheon^+$ datasets. One can observe that both the curves are well within the error bars. In both the approaches, the state finder pair has been examined and observed that it aligned with the $\Lambda$CDM model at present. The age of the Universe according to $CC$ and $Pantheon^+$ are obtained respectively as $13.81 Gyr$ and $13.96 Gyr$.
We conclude that the exponential Weyl-type $f(Q,T)$ gravity model effectively captures the evolutionary behavior of the expanding Universe and has the ability to provide accelerating behaviour at late times.\\

\begin{table}
\renewcommand\arraystretch{1.5}
\centering 
\begin{tabular}{|c|c|c|c|} 
\hline\hline 
   \parbox[c][1.5cm]{2cm}{Cosmological Parameters
    }& \parbox[c][1.5cm]{2cm}{Dynamical System
    }  &  \parbox[c][1.5cm]{2cm}{CC Dataset
    } & \parbox[c][1.5cm]{2cm}{$Pantheon^+$ Dataset
    }  \\ [0.3ex] 
\hline\hline
$q_0$ & $-0.736$ &  $-0.568$ & $-0.611$ \\[0.5ex] 
\hline 
$z_{tr}$ & $0.714$ &  $0.852$ & $0.934$ \\[0.5ex] 
\hline 
$w_{tot}$ & $-0.824$ &  $-0.712$ & $-0.740$ \\
\hline
$r_0$ & $-0.427$ &  $0.488$ & $0.583$ \\[0.5ex] 
\hline 
$s_0$ & $0.159$  &  $0.125$ & $0.384$ \\[0.5ex] 
\hline 
\end{tabular}
\caption{Cosmological parameter from Dynamical system and from Observational dataset .} \label{Table-II}

\end{table}

\section*{Acknowledgement} RB acknowledges the financial support provided by the University Grants Commission (UGC) through Junior Research Fellowship UGC-Ref. No.: 211610028858 to carry out the research work. BM acknowledges the support of IUCAA, Pune (India), through the visiting associateship program.

 \section*{Appendix}\label{Appendix}
\begin{figure}[H]
    \centering
    \includegraphics[width=8cm]{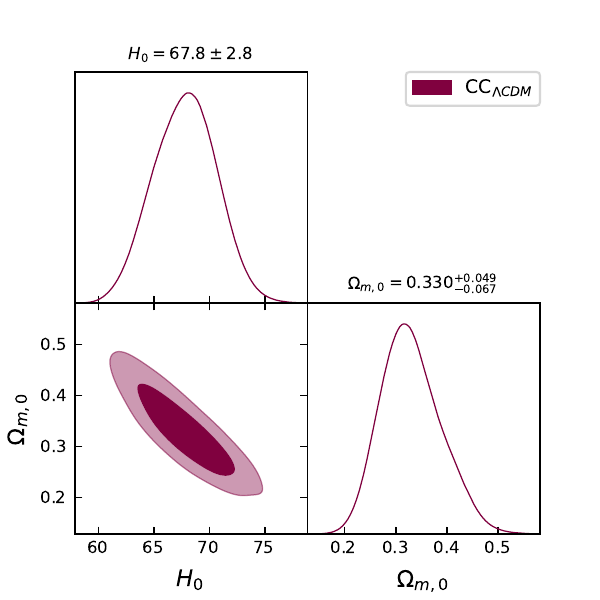}
    \caption{Contour Plot $\Lambda$CDM for $CC$ dataset} 
    \label{Fig18}
\end{figure}

\begin{figure}[H]
    \centering
    \includegraphics[width=9cm]{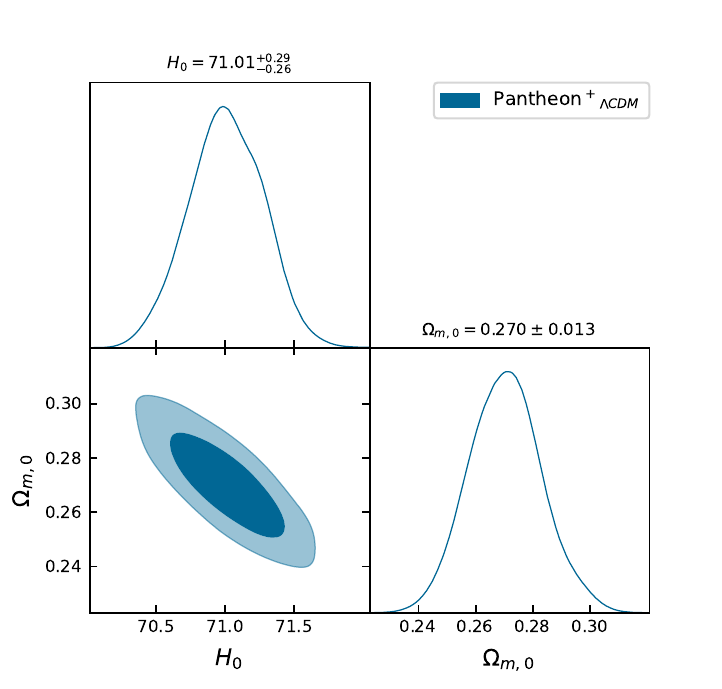}
    \caption{Contour Plot $\Lambda$CDM for $Pantheon^+$ dataset} 
    \label{Fig19}
\end{figure}

In our comparison of the exponential model against the $CC$ and $Pantheon^+$ data, we also present the corresponding \( \Lambda \)CDM MCMC results for transparency. The outcomes for \( \Lambda \)CDM are provided here for reference. FIG. \(\ref{Fig18}\) and \(\ref{Fig19}\) illustrate the MCMC posteriors and confidence regions across various priors applied to both datasets. The analysis shows rapid convergence across all prior and data combinations, resulting in nearly Gaussian uncertainties. From this, we derive \( H_0 \) and \( \Omega_{m,0} \) values of $67.80\pm2.8~kms^{-1}Mpc^{-1}$ and $0.330^{+0.049}_{-0.067}$ for the $CC$ dataset, and $71.01^{+0.29}_{-0.26}~kms^{-1}Mpc^{-1}$ and $0.270\pm0.013$ for the $Pantheon^+$ dataset. The minimum chi-squared values (\(\chi^2_{\text{min}}\)) are 14.55 for $CC$ and 1093.001 for $Pantheon^+$, respectively.


\section*{References}
\bibliographystyle{utphys}
\bibliography{references}

\providecommand{\href}[2]{#2}\begingroup\raggedright\begin{thebibliography}{10}

\bibitem{Riess_1998_116}
{\bf Supernova Search Team} Collaboration, A.~G. Riess {\em et al.}, ``{Observational evidence from supernovae for an accelerating universe and a cosmological constant},'' \href{http://dx.doi.org/10.1086/300499}{{\em Astron. J.} {\bf 116} (1998)  1009--1038}, \href{http://arxiv.org/abs/astro-ph/9805201}{{\tt arXiv:astro-ph/9805201}}.

\bibitem{Perlmutter_1998_517}
{\bf Supernova Cosmology Project} Collaboration, S.~Perlmutter {\em et al.}, ``{Measurements of $\Omega$ and $\Lambda$ from 42 high redshift supernovae},'' \href{http://dx.doi.org/10.1086/307221}{{\em Astrophys. J.} {\bf 517} (1999)  565--586}, \href{http://arxiv.org/abs/astro-ph/9812133}{{\tt arXiv:astro-ph/9812133}}.

\bibitem{Weinberg_1989_61}
S.~Weinberg, ``{The Cosmological Constant Problem},'' \href{http://dx.doi.org/10.1103/RevModPhys.61.1}{{\em Rev. Mod. Phys.} {\bf 61} (1989)  1--23}, \href{http://arxiv.org/abs/astro-ph/0005265}{{\tt arXiv:astro-ph/0005265 [astro-ph.CO]}}.

\bibitem{Appleby_2018_07}
S.~Appleby and E.~V. Linder, ``{The Well-Tempered Cosmological Constant},'' \href{http://dx.doi.org/10.1088/1475-7516/2018/07/034}{{\em J. Cosmol. Astropart. Phys.} {\bf 07} (2018)  034}, \href{http://arxiv.org/abs/1805.00470}{{\tt arXiv:1805.00470 [gr-qc]}}.

\bibitem{DiValentino_2021_38}
E.~Di~Valentino, O.~Mena, S.~Pan, L.~Visinelli, W.~Yang, A.~Melchiorri, D.~F. Mota, A.~G. Riess, and J.~Silk, ``{In the realm of the Hubble tension\textemdash{}a review of solutions},'' \href{http://dx.doi.org/10.1088/1361-6382/ac086d}{{\em Class. Quant. Grav.} {\bf 38} (2021) no.~15, 153001}, \href{http://arxiv.org/abs/2103.01183}{{\tt arXiv:2103.01183 [astro-ph.CO]}}.

\bibitem{Abdalla_2022_34}
E.~Abdalla {\em et al.}, ``{Cosmology intertwined: A review of the particle physics, astrophysics, and cosmology associated with the cosmological tensions and anomalies},'' \href{http://dx.doi.org/10.1016/j.jheap.2022.04.002}{{\em J. High Energy Astrophys.} {\bf 34} (2022)  49--211}, \href{http://arxiv.org/abs/2203.06142}{{\tt arXiv:2203.06142 [astro-ph.CO]}}.

\bibitem{Riess_2021_908}
A.~G. Riess, S.~Casertano, W.~Yuan, {\em et al.}, ``{Cosmic Distances Calibrated to 1{\%} Precision with Gaia {EDR}3 Parallaxes and Hubble Space Telescope Photometry of 75 Milky Way Cepheids Confirm Tension with $\Lambda${CDM}},'' \href{http://dx.doi.org/10.3847/2041-8213/abdbaf}{{\em Astrophys. J. Lett.} {\bf 908} (2021) no.~1, L6}.

\bibitem{Wong_2019_498}
K.~C. Wong, S.~H. Suyu, G.~C.-F. Chen, {\em et al.}, ``{H0LiCOW \textendash{} XIII. A 2.4 per cent measurement of $H_0$ from lensed quasars: 5.3\ensuremath{\sigma} tension between early- and late-Universe probes},'' \href{http://dx.doi.org/10.1093/mnras/stz3094}{{\em Mon. Not. Roy. Astron. Soc.} {\bf 498} (2020) no.~1, 1420--1439}, \href{http://arxiv.org/abs/1907.04869}{{\tt arXiv:1907.04869 [astro-ph.CO]}}.

\bibitem{Gibbons_1987}
G.~Gibbons and P.~Townsend, ``{Cosmological evolution of degenerate vacua},'' \href{http://dx.doi.org/10.1016/0550-3213(87)90700-0}{{\em Nuclear Physics B} {\bf 282} (1987) no.~-, 610}.

\bibitem{Akarsu_2021}
O.~Akarsu, S.~Kumar, E.~\"Oz\"ulker, and J.~A. Vazquez, ``{Relaxing cosmological tensions with a sign switching cosmological constant},'' \href{http://dx.doi.org/10.1103/PhysRevD.104.123512}{{\em Phys. Rev. D} {\bf 104} (2021) no.~24, 123512}, \href{http://arxiv.org/abs/2108.09239}{{\tt arXiv:2108.09239}}.

\bibitem{Riess_2004_607}
A.~G. Riess, L.-G. Strolger, J.~Tonry, {\em et al.}, ``Type {Ia} supernova discoveries at $z>1$ from the hubble space telescope: Evidence for past deceleration and constraints on dark energy evolution,'' \href{http://dx.doi.org/10.1086/383612}{{\em Astrophys. J.} {\bf 607} (2004) no.~2, 665--687}.

\bibitem{Moresco_2022_25}
M.~Moresco, L.~Amati, L.~Amendola, {\em et al.}, ``{Unveiling the Universe with emerging cosmological probes},'' \href{http://dx.doi.org/10.1007/s41114-022-00040-z}{{\em Living Rev. Rel.} {\bf 25} (2022)  6}, \href{http://arxiv.org/abs/2201.07241}{{\tt arXiv:2201.07241}}.

\bibitem{Raichoor_2020_500}
A.~Raichoor, A.~de~Mattia, A.~J. Ross, {\em et al.}, ``{The completed {SDSS}-{IV} extended Baryon Oscillation Spectroscopic Survey: large-scale structure catalogues and measurement of the isotropic {BAO} between redshift 0.6 and 1.1 for the Emission Line Galaxy Sample},'' \href{http://dx.doi.org/10.1093/mnras/staa3336}{{\em Mon. Not. Roy. Astron. Soc.} {\bf 500} (2020) no.~3, 3254--3274}.

\bibitem{Simon_2005_71}
J.~Simon, L.~Verde, and R.~Jimenez, ``{Constraints on the redshift dependence of the dark energy potential},'' \href{http://dx.doi.org/10.1103/PhysRevD.71.123001}{{\em Phys. Rev. D} {\bf 71} (2005) no.~18, 123001}.

\bibitem{Perenon_2015_11}
L.~P\'erenon, F.~Piazza, C.~Marinoni, {\em et al.}, ``Phenomenology of dark energy: general features of large-scale perturbations,'' \href{http://dx.doi.org/10.1088/1475-7516/2015/11/029}{{\em J. Cosmol. Astropart. Phys.} {\bf 2015} (2015) no.~11, 029}.

\bibitem{Jimenez_2018_039}
J.~B. Jim\'enez, L.~Heisenberg, and T.~S. Koivisto, ``{Teleparallel Palatini theories},'' \href{http://dx.doi.org/10.1088/1475-7516/2018/08/039}{{\em J. Cosmol. Astropart. Phys.} {\bf 2018} (2018) no.~08, 039}.

\bibitem{Zhao_2022_82}
D.~Zhao, ``{Covariant formulation of $f(Q)$ theory},'' \href{http://dx.doi.org/10.1140/epjc/s10052-022-10266-4}{{\em EPJC} {\bf 82} (2022)  303}.

\bibitem{Harko_2018_98}
T.~Harko, T.~S. Koivisto, F.~S.~N. Lobo, G.~J. Olmo, and D.~Rubiera-Garcia, ``{Coupling matter in modified $Q$ gravity},'' \href{http://dx.doi.org/10.1103/PhysRevD.98.084043}{{\em Phys. Rev. D} {\bf 98} (2018) no.~13, 084043}.

\bibitem{Lazkoz_2019_100}
R.~Lazkoz, F.~S.~N. Lobo, M.~Ortiz-Ba\~nos, and V.~Salzano, ``{Observational constraints of $f(Q)$ gravity},'' \href{http://dx.doi.org/10.1103/PhysRevD.100.104027}{{\em Phys. Rev. D} {\bf 100} (2019)  104027}, \href{http://arxiv.org/abs/1907.13219}{{\tt arXiv:1907.13219}}.

\bibitem{Jimenez_2020_101}
J.~B. Jim\'enez, L.~Heisenberg, T.~Koivisto, and S.~Pekar, ``{Cosmology in $f(Q)$ geometry},'' \href{http://dx.doi.org/10.1103/PhysRevD.101.103507}{{\em Phys. Rev. D} {\bf 101} (2020) no.~16, 103507}.

\bibitem{Barros_2020_30}
B.~J. Barros, T.~Barreiro, T.~Koivisto, and N.~J. Nunes, ``{Testing $f(Q)$ gravity with redshift space distortions},'' \href{http://dx.doi.org/10.1016/j.dark.2020.100616}{{\em Physics of the Dark Universe} {\bf 30} (2020)  100616}.

\bibitem{Anagnostopoulos_2021_822}
F.~K. Anagnostopoulos, S.~Basilakos, and E.~N. Saridakis, ``{First evidence that non-metricity $f(Q)$ gravity could challenge $\Lambda$CDM},'' \href{http://dx.doi.org/10.1016/j.physletb.2021.136634}{{\em Physics Letters B} {\bf 822} (2021)  136634}, \href{http://arxiv.org/abs/2104.15123}{{\tt arXiv:2104.15123}}.

\bibitem{Xu_2020_80}
Y.~Xu, T.~Harko, S.~Shahidi, and S.-D. Liang, ``{Weyl type f(Q, T) gravity, and its cosmological implications},'' \href{http://dx.doi.org/10.1140/epjc/s10052-020-8023-6}{{\em The European Physical Journal C} {\bf 80} (2020) no.~5, 708}, \href{http://arxiv.org/abs/2005.04025}{{\tt arXiv:2005.04025}}.

\bibitem{Zia_2021}
R.~Zia, D.~C. Maurya, and A.~K. Shukla, ``{Transit cosmological models in modified f(Q,T) gravity},'' \href{http://dx.doi.org/10.1142/S0219887821500511}{{\em International Journal of Geometric Methods in Modern Physics} {\bf 18} (2021) no.~04, 2150051}.

\bibitem{Sharif_2024}
M.~Sharif and I.~Ibrar, ``{Analysis of reconstructed modified f(Q,T) gravity},'' \href{http://dx.doi.org/10.1016/j.cjph.2024.04.026}{{\em Physics Letters B} {\bf 89} (2024) no.~-, 1578--1594}.

\bibitem{Narawade_2023_992}
S.~Narawade, M.~Koussour, and B.~Mishra, ``{Constrained f(Q,T) gravity accelerating cosmological model and its dynamical system analysis},'' \href{http://dx.doi.org/10.1016/j.nuclphysb.2023.116233}{{\em Nuclear Physics B} {\bf 992} (2023) no.~-, 116233}, \href{http://arxiv.org/abs/2305.08145}{{\tt arXiv:2305.08145}}.

\bibitem{Pati_2021}
L.~Pati, B.~Mishra, and S.~K. Tripathy, ``{Model parameters in the context of late time cosmic acceleration in f(Q,T) gravity},'' \href{http://dx.doi.org/10.1088/1402-4896/ac0f92}{{\em Physica Scripta} {\bf 96} (2021) no.~105003, 1578--1594}.

\bibitem{Najera_2021_34}
A.~N\'ajera and A.~Fajardo, ``{Fitting f(Q,T) gravity models with a $\Lambda$CDM limit using H(z) and Pantheon data},'' \href{http://dx.doi.org/10.1016/j.dark.2021.100889}{{\em Physics of the Dark Universe} {\bf 34} (2021) no.~-, 100889}, \href{http://arxiv.org/abs/2104.14065}{{\tt arXiv:2104.14065}}.

\bibitem{Bhagat_2023_42}
R.~Bhagat, S.~Narawade, B.~Mishra, and S.~Tripathy, ``{Constrained cosmological model in f(Q,T) gravity with non-linear non-metricity},'' \href{http://dx.doi.org/10.1016/j.dark.2023.101358}{{\em Physics of the Dark Universe} {\bf 42} (2023) no.~-, 101358}, \href{http://arxiv.org/abs/2308.11190}{{\tt arXiv:2308.11190}}.

\bibitem{Narawade_2024Baryon}
S.~A. Narawade, S.~K. Tripathy, R.~Patra, and B.~Mishra, ``{Baryon Asymmetry Constraints on Extended Symmetric Teleparallel Gravity},'' \href{http://dx.doi.org/10.1134/s0202289324700026}{{\em Gravitation and Cosmology} {\bf 30} (2024) no.~-, 135}, \href{http://arxiv.org/abs/2402.16275}{{\tt arXiv:2402.16275}}.

\bibitem{Wheeler_2014}
J.~T. Wheeler, ``{Weyl gravity as general relativity},'' \href{http://dx.doi.org/10.1103/PhysRevD.90.025027}{{\em Phys. Rev. D} {\bf 90} (2014) no.~7, 025027}.

\bibitem{Alvarez_2017}
E.~Alvarez and S.~Gonzalez-Martin, ``{Weyl gravity revisited},'' \href{http://dx.doi.org/10.1088/1475-7516/2017/02/011}{{\em Journal of Cosmology and Astroparticle Physics} {\bf 2017} (2017) no.~02, 011}, \href{http://arxiv.org/abs/1610.03539}{{\tt arXiv:1610.03539}}.

\bibitem{Gomes_2019}
C.~Gomes and O.~Bertolami, ``{Nonminimally coupled Weyl gravity},'' \href{http://dx.doi.org/10.1088/1361-6382/ab52b9}{{\em Classical and Quantum Gravity} {\bf 36} (2019) no.~23, 235016}, \href{http://arxiv.org/abs/1812.04976}{{\tt arXiv:1812.04976}}.

\bibitem{Haghani_2013_88}
Z.~Haghani, T.~Harko, H.~R. Sepangi, and S.~Shahidi, ``{Weyl-Cartan-Weitzenb$\ddot{o}$ck gravity through Lagrange multiplier},'' \href{http://dx.doi.org/10.1103/physrevd.88.044024}{{\em Phys. Rev. D} {\bf 88} (2013) no.~4, }, \href{http://arxiv.org/abs/1307.2229}{{\tt arXiv:1307.2229 [gr-qc]}}.

\bibitem{Yang_2021_81}
J.-Z. Yang, S.~Shahidi, T.~Harko, and S.-D. Liang, ``{Geodesic deviation, Raychaudhuri equation, Newtonian limit, and tidal forces in Weyl-type f(Q, T) gravity},'' \href{http://dx.doi.org/10.1140/epjc/s10052-021-08910-6}{{\em The European Physical Journal C} {\bf 81} (2021) no.~2, 111}, \href{http://arxiv.org/abs/2101.09956}{{\tt arXiv:2101.09956}}.

\bibitem{Bhagat_2023_41}
R.~Bhagat, S.~Narawade, and B.~Mishra, ``{Weyl type f(Q,T) gravity observational constrained cosmological models},'' \href{http://dx.doi.org/10.1016/j.dark.2023.101250}{{\em Physics of the Dark Universe} {\bf 41} (2023)  101250}, \href{http://arxiv.org/abs/2305.01659}{{\tt arXiv:2305.01659}}.

\bibitem{Zhadyranova_2024}
A.~Zhadyranova, M.~Koussour, and S.~Bekkhozhayev, ``{The dynamics of matter bounce cosmology in Weyl$-$type f(Q, T) gravity},'' \href{http://dx.doi.org/10.1016/j.cjph.2024.04.023}{{\em Chinese Journal of Physics} {\bf 89} (2024) no.~-, 1483}, \href{http://arxiv.org/abs/2406.15409}{{\tt arXiv:2406.15409}}.

\bibitem{Bhagat_ASPdyna2024}
R.~Bhagat and B.~Mishra, ``{Observational constrained Weyl type f(Q,T) gravity cosmological model and the dynamical system analysis},'' \href{http://dx.doi.org/10.1016/j.astropartphys.2024.103011}{{\em Astroparticle Physics} {\bf 163} (2024) no.~-, 103011}, \href{http://arxiv.org/abs/2402.18915}{{\tt arXiv:2402.18915}}.

\bibitem{Paliathanasis_2023_41}
A.~Paliathanasis, ``{Dynamical analysis of f(Q) cosmology},'' \href{http://dx.doi.org/10.1016/j.dark.2023.101255}{{\em Physics of the Dark Universe} {\bf 41} (2023) no.~-, 101255}, \href{http://arxiv.org/abs/2304.04219v2}{{\tt arXiv:2304.04219v2}}.

\bibitem{Bohmer_2016_book_dyna}
C.~G. B\"{o}hmer and N.~Chan, ``{Dynamical Systems in Cosmology},'' \href{http://dx.doi.org/10.1142/9781786341044_0004}{{\em Dynamical and Complex Systems} {\bf -} (2017) no.~-, 121--156}, \href{http://arxiv.org/abs/1409.5585v2}{{\tt arXiv:1409.5585v2}}.

\bibitem{Pati_2023_83}
L.~Pati, S.~A. Narawade, S.~K. Tripathy, and B.~Mishra, ``{Evolutionary behaviour of cosmological parameters with dynamical system analysis in f(Q, T) gravity},'' \href{http://dx.doi.org/10.1140/epjc/s10052-023-11598-5}{{\em The European Physical Journal C} {\bf 83} (2023) no.~5, 445}, \href{http://arxiv.org/abs/2206.11928}{{\tt arXiv:2206.11928}}.

\bibitem{Lohakare_2023_39}
S.~V. Lohakare, B.~Mishra, S.~Maurya, {\em et al.}, ``{Analyzing the geometrical and dynamical parameters of modified Teleparallel-Gauss{\textendash}Bonnet model},'' \href{http://dx.doi.org/10.1016/j.dark.2022.101164}{{\em Phys. Dark Univ.} {\bf 39} (2023)  101164}, \href{http://arxiv.org/abs/2209.13197}{{\tt arXiv:2209.13197 [gr-qc]}}.

\bibitem{Lu_2019_79}
J.~Lu, X.~Zhao, and G.~Chee, ``{Cosmology in symmetric teleparallel gravity and its dynamical system},'' \href{http://dx.doi.org/10.1140/epjc/s10052-019-7038-3}{{\em The European Physical Journal C} {\bf 79} (2019)  530}.

\bibitem{Duchaniya_2023_83}
L.~K. Duchaniya, S.~A. Kadam, J.~L. Said, {\em et al.}, ``{Dynamical systems analysis in $f(T, \phi)$ gravity},'' \href{http://dx.doi.org/10.1140/epjc/s10052-022-11155-6}{{\em Eur. Phys. J. C} {\bf 83} (2023) no.~1, }, \href{http://arxiv.org/abs/2209.03414}{{\tt arXiv:2209.03414 [gr-qc]}}.

\bibitem{Coley_1999}
A.~A. Coley, ``{Dynamical systems in cosmology},'' in {\em Spanish Relativity Meeting (ERE 99)}.
\newblock 9, 1999.
\newblock \href{http://arxiv.org/abs/gr-qc/9910074}{{\tt arXiv:gr-qc/9910074}}.

\bibitem{aulbach_1984_1058}
B.~Aulbach, ``Continuous and discrete dynamics near manifolds of equilibria,'' {\em Lecture notes in mathematics} {\bf 1058} (1984)  .

\bibitem{Kadam_2022_82}
S.~A. Kadam, B.~Mishra, and J.~Said~Levi, ``{Teleparallel scalar-tensor gravity through cosmological dynamical systems},'' \href{http://dx.doi.org/10.1140/epjc/s10052-022-10648-8}{{\em Eur. Phys. J. C} {\bf 82} (2022) no.~8, 680}, \href{http://arxiv.org/abs/2205.04231}{{\tt arXiv:2205.04231 [gr-qc]}}.

\bibitem{Bertone_2004_405}
G.~Bertone, D.~Hooper, and J.~Silk, ``{Particle dark matter: Evidence, candidates and constraints},'' \href{http://dx.doi.org/10.1016/j.physrep.2004.08.031}{{\em Phys. Rept.} {\bf 405} (2005)  279--390}, \href{http://arxiv.org/abs/hep-ph/0404175}{{\tt arXiv:hep-ph/0404175}}.

\bibitem{Sahni_2003_77}
V.~Sahni, T.~Saini, A.~Starobinsky, and U.~Alam, ``{Statefinder--A new geometrical diagnostic of dark energy},'' \href{http://dx.doi.org/10.1134/1.1574831}{{\em JETP Letters} {\bf 77} (2003) no.~5, 201 -- 206}, \href{http://arxiv.org/abs/astro-ph/0201498}{{\tt arXiv:astro-ph/0201498}}.

\bibitem{Alam_2003}
U.~Alam, V.~Sahni, T.~Deep~Saini, and A.~A. Starobinsky, ``{Exploring the expanding Universe and dark energy using the statefinder diagnostic},'' \href{http://dx.doi.org/10.1046/j.1365-8711.2003.06871.x}{{\em Monthly Notices of the Royal Astronomical Society} {\bf 344} (2003) no.~4, 1057}, \href{http://arxiv.org/abs/astro-ph/0303009}{{\tt arXiv:astro-ph/0303009}}.

\bibitem{Foreman-Mackey_2013_125}
D.~Foreman-Mackey, D.~W. Hogg, D.~Lang, and J.~Goodman, ``{emcee: The MCMC Hammer},'' \href{http://dx.doi.org/10.1086/670067}{{\em Publications of the Astronomical Society of the Pacific} {\bf 125} (2013) no.~925, 306}, \href{http://arxiv.org/abs/1202.3665}{{\tt arXiv:1202.3665}}.

\bibitem{Rebecca_2022}
R.~Briffa, C.~Escamilla-Rivera, J.~L. Said, and other, ``{Impact of $H_0$ priors on f(T) late time cosmology },'' \href{http://dx.doi.org/10.1140/epjp/s13360-022-02725-4}{{\em The European Physical Journal Plus} {\bf 137} (2022) no.~5, 532}, \href{http://arxiv.org/abs/2108.03853}{{\tt arXiv:2108.03853}}.

\bibitem{Cowan_2002}
J.~J. Cowan, C.~Sneden, S.~Burles, I.~I. Ivans, T.~C. Beers, J.~W. Truran, J.~E. Lawler, F.~Primas, G.~M. Fuller, B.~Pfeiffer, and K.~Kratz, ``{The Chemical Composition and Age of the Metal$-$poor Halo Star BD $+17o3248$},'' \href{http://dx.doi.org/10.1086/340347}{{\em The Astrophysical Journal} {\bf 572} (2002) no.~2, 861}, \href{http://arxiv.org/abs/0202429}{{\tt arXiv:0202429}}.

\end{thebibliography}\endgroup

\end{document}